\def\dend#1{{\if*#1{\it Paenibacillus dendritiformis}\else
                {\it P. dendritiformis}\fi}}
\def\Tvar{var. {\it dendron}}
\def\Cvar{var. {\it chiralis}}
\def\Tname#1{{\if*#1\dend* \Tvar\else
                \if-#1\dend{} \Tvar\else
                 \dend{} \Tvar{} #1\fi\fi}}
\def\Cname#1{{\if*#1\dend* \Cvar\else
                \if-#1\dend{} \Cvar\else
                 \dend{} \Cvar{} #1\fi\fi}}
\def\Vname#1{{\if*#1\eddi{Paenibacillus}{V}\else
                \if-#1\eddi{P.}{V}\else \eddi{P.}{V} #1\fi\fi}}
\def\bsub#1{{\if*#1{\it Bacillus subtilis}\else
                \if-#1{\it B. subtilis}\else {\it B. subtilis} #1\fi\fi}}
\def\bacil#1{\if *#1{Bacillus}\else{B.}\fi}
\def\bcirc#1{\if *#1{\it \bacil* circulans}\else
                \if -#1{\it \bacil{} circulans}\else
                   {\it \bacil{} circulans} #1\fi\fi}
\def\ecoli#1{{\if*#1{\it Escherichia coli}\else
                \if-#1{\it E. coli}\else {\it E. coli} #1\fi\fi}}
\def\salmon#1{{\if*#1{\it Salmonella typhimurium}\else
                \if-#1{\it S. typhimurium}\else
                {\it S. typhimurium} #1\fi\fi}}

\def\myxo#1{{\if*#1{\it Myxococcus xanthus}\else
                \if-#1{\it M. xanthus}\else
                {\it M. xanthus} #1\fi\fi}}

\def\T{{${\cal T }$} }

\def\be{\begin{equation}}
\def\ee{\end{equation}}
\def\ben{\begin{enumerate}}
\def\een{\end{enumerate}}
\def\ba{\begin{eqnarray}}
\def\ea{\end{eqnarray}}
\def\partderiv#1#2{{\partial #1\over\partial #2}}

\def\Tm{{\T morphotype }}

\def\Tme#1{{\T morphotype#1}}

\def\etal{{\it et al. }}

\def\text#1{\hbox{#1}}

\documentstyle[pre,aps,preprint,epsf]{revtex}

\begin{document}
\draft

\title{Studies of Bacterial Branching Growth using Reaction-Diffusion
  Models for Colonial Development}
\author{Ido Golding, Yonathan Kozlovsky, Inon Cohen and Eshel Ben-Jacob}
\address{School of Physics and Astronomy, Raymond and Beverly Sackler
  Faculty of Exact Sciences, \\ Tel Aviv University,
Tel Aviv 69 978, Israel}
\maketitle


\begin{abstract}
Various bacterial strains exhibit colonial branching patterns during
growth on poor substrates. These patterns reflect bacterial
cooperative self-organization and cybernetic processes of
communication, regulation and control employed during colonial
development.
One method of modeling is the continuous, or coupled reaction-diffusion
approach, in which continuous time evolution equations describe the
bacterial density and the concentration of the relevant chemical
fields. In the context of branching growth, this idea has been pursued
by a number of groups. We present an additional model which
includes a lubrication fluid excreted by the bacteria.
We also  add fields of chemotactic agents to the other
models. We then present a critique of this whole enterprise with
focus on the models' potential for revealing new biological features.
\end{abstract}



\section{Introduction}

During the course of evolution, bacteria 
have developed sophisticated
cooperative behavior and intricate communication capabilities  \cite{SDA57,Shap88,BenJacob97,BCL98,LB98}. These include: direct cell-cell physical interactions via extra-membrane
polymers \cite{Mend78,Devreotes89}, collective production of
extracellular "wetting" fluid for movement on hard surfaces
\cite{MKNIHY92,Harshey94}, long range chemical signaling, such as
quorum sensing \cite{FWG94,LWFBSLW95,FWG96} and chemotactic signaling\footnote{
  Chemotaxis is a bias of
movement according to the gradient of a chemical agent. Chemotactic
signaling is a chemotactic response to an agent emitted by the bacteria.}
\cite{BB91,BE95,BB95}, collective activation and deactivation of genes
\cite{ST91,SM93,MS96} and even exchange of genetic material
\cite{GR95,RPF95,Miller98}. Utilizing these capabilities, bacterial colonies
develop complex spatio-temporal patterns in response to adverse growth
conditions.

It is now understood that the study of cooperative self-organization
of bacterial colonies is an exciting new multidisciplinary field of
research, necessitating the merger of biological information with the
physics of non-equilibrium processes and the mathematics of
non-linear dynamics.
At this stage, several experimental systems have been identified, and
preliminary modeling efforts are making significant progress in
providing a framework for the understanding of experimental
observations \cite{MKNIHY92,MS96,Kessler85,FM89,PK92a,BSST92,MHM93%
,BSTCCV94a,BCSALT95,WTMMBB95,BCCVG97,KW97,ES98}.

Fujikawa and Matsushita \cite{FM89,MF90,FM91} reported for the
first time \footnote{
We refer to the first time that branching growth was studied as
such. Observations of branching colonies occurred long ago
\protect\cite{SC38,Henrici48}.
} that bacterial colonies could grow elaborate branching patterns of
the type known from the study of fractal formation in the process
of diffusion-limited-aggregation (DLA) \cite{WitSan81,Sander86,Vicsek89} . This
work was done with \bsub*, but was subsequently extended to other
bacterial species such as {\it Serratia marcescens} and {\it
Salmonella anatum} \cite{MM95}. It was shown explicitly that nutrient
diffusion was the relevant dynamics responsible for the growth instability.
Later, we will see how models which couple nutrient diffusion to
bacterial density can naturally account for these structures.

Motivated by these observations, Ben-Jacob \etal \cite{BSST92,BTSA94,BSTCCV94a}
conducted new experiments to see how adaptive
bacterial colonies could be in the presence of external ``pressure'',
here in the form of a limited nutrient supply and hard surface.
The endeavor started with \bsub 168, which is non-motile on a solid agar
surface, from which a new species of bacteria has been isolated
\cite{BSST92,BTSA94}.
The new species was designated \Tname* \cite{TBG98}.
This species is motile on the hard surface and its colonies
exhibit branching patterns (Fig. 1).
The new mode of tip-splitting growth was found to be inheritable and
transferable by a single cell, hence it is
referred to as a distinctive morphotype \cite{BCG98}, and, to indicate the
tip-splitting character of the growth, it was denoted \Tme.
In the next section we describe in some detail the observations of
Ben-Jacob \etal and
Matsushita \etal. Additional studies of branching
colonial growth are reported by Matsuyama \etal \cite{MKNIHY92,MM93}
and
Mendelson and Salhi \cite{MS96}.

All the various strains reported in the studies quoted above exhibit
branching patterns during growth on a poor substrate.
Drawing
on the analogy with diffusive patterning in non-living systems
\cite{KKL88,Langer89,BG90,BenJacob93}, we can state that 
complex patterns are expected.  The cellular reproduction
rate that determines the growth rate of the colony is limited by the
level of nutrients available for the cells. The latter is limited by
the diffusion of nutrients towards the colony (for low nutrient
substrate). Hence colony growth under certain conditions should be similar to diffusion
limited growth in non-living systems as mentioned above
\cite{BG90,BenJacob93}.  The study of diffusive
patterning in non-living systems teaches us that the diffusion field
drives the system towards decorated (on many length scales) irregular
fractal shapes.  Indeed, bacterial
colonies can develop patterns reminiscent of those observed during
growth in non-living systems.
But, this is certainly not the end of the story. The colonies exhibit
a richer behavior. This, ultimately, is a reflection of the additional
levels of complexity involved when the building blocks of the colonies,
the bacteria, are themselves living systems. We now start to reveal
the cybernetic processes (communication, regulation and control)
which are part of the colonial adaptive self-organization, and their
determination of the interaction between genetic information and
biophysical behavior.

How should one approach the modeling of the complex bacterial patterning?
With present computational power it is natural to use computer models as a
main tool in the study of complex systems. However, one must be careful not
to be trapped in the "reminiscence syndrome", described by J. D. Cowan
\cite{Horgan95}, as the tendency to devise a set of rules which will
mimic some aspect of the observed phenomena and then, to quote J. D.
Cowan "They say: `Look, isn't this reminiscent of a biological or
physical phenomenon!' They jump in right away as if it's a decent
model for the phenomenon, and usually of course it's just got some
accidental features that make it look like something." Yet the
reminiscence modeling approach has some indirect value.  True, doing
so does not reveal (directly) the biological functions and behavior.
However, it does reflect understanding of geometrical and temporal
features of the patterns, which indirectly might help in revealing
the underlying biological principles. Another extreme is the
"realistic modeling" approach, where one constructs an algorithm that
includes in details all the known biological facts about the system.
Such an approach sets a trajectory of ever including more and more
details (vs. generalized features). The model keeps evolving to
include so many details that it loses any predictive power.

Here we try to promote another approach -- the "generic modeling" one
\cite{KL93,BSTCCV94a,Azbel93,KW97}. We seek to elicit, from the
experimental observations and the biological knowledge, the generic
features and basic principles needed to explain the biological
behavior and to include these features in the model. We will
demonstrate that such modeling, with close comparison to experimental
observations, can be used as a research tool to reveal new
understanding of the biological systems.

Generic modeling is not about using sophisticated, as it may, mathematical
description to dress pre-existing understanding of complex biological
behavior. Rather, it means a cooperative approach, using existing biological
knowledge together with mathematical tools and synergetic point of view for
complex systems to reach a new understanding (which is reflected in the
constructed model) of the observed complex phenomena.

The generic models can yet be grouped into two main categories:
1. Discrete models such as the communicating walkers models of
Ben-Jacob \etal \cite{BSTCCV94a,BCSCV95,BCCVG97} and the bions model
of Kessler and Levine \cite{KL93,KLT97}.
In this approach, the microorganisms (bacteria in the first model and
amoebae in second) are represented by discrete, random walking
entities (walkers and bions, respectively) which can consume
nutrients, reproduce, perform random or biased movement, and produce
or respond to chemicals. The time evolution of the chemicals is
described by reaction-diffusion equations.
2. Continuous or reaction-diffusion models \cite{PS78,Mackay78}. In
these models the microorganisms are represented via their 2D density,
and a reaction-diffusion equation of this density describes their
time evolution. This equation is coupled to the other
reaction-diffusion equations for the chemical fields. In the context
of branching growth, this idea has been pursued recently by Mimura and
Matsushita
\etal \cite{Mimura97,MWIRMSM98},
Kawasaki \etal \cite{KMMUS97} and Kitsunezaki \cite{Kitsunezaki97}. A
summary and critique of this approach is provided by Rafols
\cite{Rafols98}.

Here we describe a new model which includes a lubrication fluid and a
model with a cutoff, as was proposed by Kessler and Levine
\cite{KL98}.
We compare the results obtained by the various models and the
experimental observations. Our main goal is to identify the biological
and mathematical requirements for branching patterns. We then study
the effect of nutrients- and signaling- chemotaxis, and conclude 
that chemotaxis is needed to explain the wealth of experimental observations.

\section{Experimental Results}

\label{sec:experiments}

Several strains of bacteria were reported to produce
tip-splitting branched patterns under conditions of low level of
nutrient.
We describe here the experimental results of 
Ben-Jacob {\it et al.}
\cite{BSST92,BTSA94,BSTCCV94a,BSTCCV94b} 
-- working with colonies of \Tname* (\T
morphotype) -- and Matsushita, Fujikawa, Matsuyama and coworkers \cite{MF90,FM89,FM91,MHM93,MM95}
-- working with colonies
of \bsub*.

\subsection{Growth patterns of {${\cal T }$} morphotype}

\label{sec:growth}

\setcounter{paragraph}{0}
\paragraph{Macroscopic observations}

All manner of patterns are exhibited by {${\cal T }$} morphotype as the
growth conditions are varied. An example of branching pattern is shown
in figure \ref{fig:T:example}.
This kaleidoscope of shapes may be grouped
into a number of "essential" patterns.
For intermediate agar concentrations (about 1.5\% -- $1.5g$ in
$100ml$), at very high peptone levels (above 10$g/l$) the patterns are
compact (Fig. \ref{fig:T:morphology}a).
At somewhat lower but still high peptone levels (about 5-10$g/l$)
the patterns, reminiscent of viscous fingering patterns in Hele-Shaw
devices \cite{BG90}, exhibit quite
pronounced radial symmetry and may be characterized as dense fingers
(Fig. \ref{fig:T:morphology}b), each finger being much wider than the
distance between fingers.
For intermediate peptone levels, branching patterns with lower
fractal dimension (reminiscent of electro-chemical deposition) are
observed (Fig. \ref{fig:T:morphology}c). The patterns are "bushy",
with branch width smaller than the distance between branches.
As the peptone level is lowered, the patterns become more ramified
and fractal--like. Surprisingly, at even lower peptone
levels (below 0.25$g/l$ for 2\% agar concentration) the colonies
revert to organized structures: fine branches forming a well
defined global envelope. We characterize these patterns as fine
radial branches (Fig. \ref{fig:T:morphology}d). For extremely low
peptone levels (below 0.1$g/l$), the colonies lose the fine radial
structure and again exhibit fractal patterns (Fig.
\ref{fig:T:extremes}a).  For high agar concentration the branches are
very thin (Fig. \ref{fig:T:extremes}b).

At high agar concentration and intermediate peptone levels the
colonies display a structure of concentric rings superimposed on a
branching pattern (Fig. \ref{fig:T:fricks}a).
At high agar concentration and very high peptone levels the colonies
display a structure of concentric rings in a compact colony (Fig.
\ref{fig:T:fricks}b).  At high agar concentrations the branches also
exhibit a global twist with the same handedness, as shown in figure
\ref{fig:T:fricks}c.  Similar observations during growth of other
bacterial strains have been reported by Matsuyama \etal
\cite{MM93,MHM93}. We referred to such growth patterns as having weak
chirality \cite{BCSCV95,BenJacob97}.

A closer look at an individual branch (Fig. \ref{fig:T:fricks}d)
reveals a phenomenon of density variations within the branches. These
3-dimensional structures arise from accumulation of cells in layers.
The aggregates can form spots and ridges which are either scattered
randomly, ordered in rows, or organized in a leaf-veins-like
structure.
The aggregates are not frozen; the cells in them are motile and the
aggregates are dynamically maintained.

At the other extreme, of very soft agar (0.5\% and below), the \Tm
does not exhibit branching patterns. Instead, the growth is compact
with density.
In the range of 0.5\%-1\% agar concentration the colonies typically
have a shape of many arms stars.

\paragraph{Microscopic observations}

Under the microscope, bacterial cells are seen to perform a random-walk-like
movement in a layer of fluid on the agar surface.
This wetting fluid is assumed to be excreted by the
cells and/or drawn by the cells from the agar \cite{BSTCCV94a,BSTCCV94b}.
The cellular movement is confined to this fluid; isolated cells spotted on
the agar surface do not move. This is an important observation as we
discuss later when formulating the models.
The fluid's boundary thus defines a local
boundary for the branch. Whenever the cells are active, the boundary
propagates slowly, as a result of the cellular movement and production of
wetting fluid.

At very low agar concentrations (below 0.5\%) the bacteria swim
inside the agar and not on its surface. Between 0.5\% and 1\% agar
concentration some of the bacteria move on the surface and some inside
the agar.

The observations reveal also that the cells are active at the outer
parts of the colony, while closer to the center the cells are
stationary (do not move) and some of them sporulate (form spores). It
is known that certain bacteria respond to adverse growth conditions by
entering a spore stage until more favorable growth conditions return.
Such spores are metabolically inert and exhibit a marked resistance to
the lethal effects if heat, drying, freezing, deleterious chemicals,
and radiation.

\subsection{Morphology selection, morphology diagram and velocity-pattern
correlations}

The emerging understanding of pattern determination in non-living
includes the concepts of morphology diagram, morphology selection,
morphology velocity correlations and morphology transitions
\cite{BenJacob93}.
In short, the patterns formed in many evolving azoic (non
living) systems may often be grouped into a small number of "essential
shapes" or morphologies each representing a dominance of a different
underlying effect. If each morphology is observed over a range of
growth conditions, a morphology diagram may exist. The existence of a
morphology diagram implies the existence of a morphology selection
principle and vice versa. 
Ben-Jacob \etal proposed the existence of a new morphology selection
principle: the principle of the fastest growing morphology
\cite{BJGMG88,BG90}, a principle which should be applicable for a wide
range of growth conditions.  In general, if more than one morphology
is a possible solution, only the fastest growing one is nonlinearly
stable and will be observed, that is, selected. 

The new selection principle implies that the average velocity is an
appropriate response function for describing the growth processes and
hence should be correlated with the geometrical character of the
growth. In other words, for each regime (essential shape) in the
morphology diagram, there is a characteristic functional dependence of
the velocity on the growth parameters. At the boundaries between the
regimes there is either discontinuity in the velocity (first
order-like transition) or in its slope (second order-like
transition).

At present, there is some evidence for the existence of the new
selection principle in non-living systems. 
The new principle might also be valid for pattern determination
during colonial development in bacteria \cite{BTSA94,BenJacob97}. The
bacterial patterns may be grouped into a small number of "essential
shapes", each observed over a range of growth conditions
\cite{FM89,FM91,BSST92,BTSA94,MS96}.  To prove this hypothesis, the
next step would be to demonstrate the velocity-pattern correlation
during colonial growth.

A plot of the growth velocity as a function of
nutrient level for 1.5\% agar concentration is shown in Ref.
\cite{BSTCCV94b}. For the presented range of peptone levels
it was found that the velocity shows three distinct regimes of response,
each corresponding also to a distinct morphology (the fine radial
branches, branching patterns and dense fingers), as was predicted for
non-living systems. The change in velocity suggests that the switching
between morphologies is indeed a real morphology transition and not a
simple cross-over (see Ref. \cite{BenJacob93}). The transition at low
peptone level (between fine radial branches and branching structure)
might be a first order morphology transition, i.e. a transition
characterized by a jump in the velocity and hysteresis. The transition
at the higher peptone level (from branches to dense
fingering) seems to be second-order-like. These observations of
velocity-pattern correlations strongly support the existence of a
morphology selection principle which determines the selected
colonial morphology for a given morphotype.

In non-evolving (equilibrium) systems there is a phenomenon of
critical fluctuations when the system is kept at the transition point
between two phases. At that point the system consists of a mixture of
the two phases. 
In Ref. \cite{BenJacob93} it was shown that an analogous phenomenon
exists in evolving
non-living systems  and explained that this fact
provides additional support for the idea of morphology transitions. 
Figure \ref{fig:T:transition} shows patterns exhibited by colonies
grown at "critical" peptone levels, where transitions between two
morphologies occur.
Similarly, for the fluctuations displayed by non-living systems, we
observe a combination of the morphologies characterizing the patterns
above and below the critical point. These observations provide
additional support for the relevance of the concepts of morphology
selection and morphology transition to colonial development.

\subsection{Growth patterns of \bsub{}}

Matsushita and coworkers \cite{MF90,FM89,FM91,MHM93,MM95} studied the colonial branching
patterns and morphology diagram of the bacteria specie \bsub OG-01. A
detailed summary of these observations is provided by Rafols
\cite{Rafols98}. A typical morphology diagram is shown in figure
\ref{fig:bsub:morphology}.
Note that here the x-axis is the inverse agar concentration and the y-axis
is the nutrient level. These bacteria are not efficient in producing a
lubricating fluid, hence above about $0.8 \%$ agar concentration they
can not move on the agar surface: Under such conditions and low level
of nutrients (below 1 g/l peptone), DLA-like patterns are observed.
As the level of nutrients is increased, the patterns become compact,
with a cellular structure at the interface.

For low agar concentrations (below 0.5\%, so that the bacteria can
move) and low level
of nutrients, dense branching patterns are observed. These patterns
are replaced by compact growth for higher levels of nutrients.
Beautiful patterns of concentric rings imposed on a dense branching
growth (Fig. \ref{fig:bsub:morphology}) are observed at high levels of
nutrients and intermediate agar concentric (about $0.75 \%$).
For more details see Ref \cite{Rafols98}.

The different morphologies correspond to different growth velocities;
DLA-like patterns grow in about a month, compact patterns at
intermediate concentrations of agar grow in about a week, dense
branching patterns and patterns of concentric rings grow in few days,
and compact patterns at low concentrations of agar grow in half a day.
From this we learn that indeed the growth velocity of the various
morphologies is very different. Moreover, it seems that different
bacterial movement mechanisms correspond to the different regimes. Thus we
expect a real transition between the various regimes in the morphology
diagram rather than a simple cross-over. Therefore, if velocity
as function of the growth parameters is to be measured, it will
probably show a jump in the velocity or its slope (first or second
morphology transitions, respectively).

\section{Biological Background}

\label{sec:bio-background}

Clearly we cannot begin to
encompass all the biological background. Thus we will describe here, based
on our previous experience, only the most relevant information
for the understanding and modeling of the observed colonial patterning.

\subsection{Bacterial Surface Translocation}

\label{sec:movement}

The most widely studied 
mechanism used by bacteria for movement is swimming with  flagella \cite{Eis90},
but other mechanisms exist as well \cite{Henri72}.
Most common types of bacterial movements are categorized to be
\\ $\bullet$
Swimming -- Surface translocation produced through the action of
flagella. The cells move individually and at random in the same manner
as flagellated bacteria move in wet mounts (i.e., nearly straight runs
separated by brief tumbling). Swimming takes place only in
sufficiently thick surface fluid. Microscope observations reveal no
organized flow-field pattern.
\\ $\bullet$
Swarming -- Surface translocation produced through the action
of flagella, but unlike swimming, the movement is continuous and
regularly follows the long axis of the cell. The cells are
predominantly aggregated in bundles during their movement, and
microscope observations reveal  flow-field patterns highly organized
in whirls and bands.
\\ $\bullet$
Gliding -- Surface translocation occurring only in
non-flagellated bacteria and only when in contact with solid surface.
In all other respects, gliding is identical to swarming.

\subsection{Modeling Bacterial Movement}
\label{sec:swimming}

As for the movement of \Tme, based on microscope observations of
movement and electron
microscope observations of flagella we identify the movement as
swimming.
Cells tumble about every $\tau_T\approx 1-5$ depending on external
conditions. The speed of the bacterium between tumbling events is very
sensitive to conditions such as the liquid viscosity, temperature and
pH level. Typically, it is of the order of 1-10$\mu
m/sec$.

Swimming can be approximated by a random walk with variable step size
\cite{Berg93}.
At low bacterial densities the random walk can be described by a
diffusion equation with a 
diffusion coefficient $D_b\equiv v^2\,\tau
_T=10^{-8}-10^{-5}cm^2/sec$.
Low bacterial densities means that the mean free path between
bacterial collisions $l_c$ is longer than the tumbling length
$l_T\equiv v\tau _T $, thus
collisions between the bacteria can be neglected. The mean free path
(or collision length) is
\begin{equation}
l_c\propto \left\{
\begin{array}{ll}
\rho ^{-\frac{{1}}3} & \hbox{in 3 dimensions} \\
\sigma ^{-\frac{{1}}2} & \hbox{in 2 dimensions}
\end{array}
\right.
\end{equation}
where $\rho $ is the 3D bacterial density and $\sigma $ is the 2D
density -- the projection of $\rho $ on the surface.

At high densities ($l_c<l_T$), the collisions cannot be
neglected.  In attempt to approximate the dynamics in those
conditions, one may want to consider the time of straight motion to be
$l_c/v$ instead of $\tau _T$.
Hence $D_b$ depends on the bacterial density to yield
\begin{equation}
D_b\propto \left\{
\begin{array}{ll}
v\rho ^{-\frac{{1}}3} & \hbox{in 3D} \\
v\sigma ^{-\frac{{1}}2} & \hbox{in 2D}
\end{array}
\right. ~~.  \label{eq:hige_density_diffusion:0}
\end{equation}
This approximation is valid under the assumptions that a collision
event is identical to a tumbling event (abrupt uncorrelated change in
direction of motion), that a tumbling event is independent of the
collisions, and that the speed between such events is not affected by
their frequency.

The assumption that a collision event is like a tumbling event poses
many problems. Even if the bacteria do not activate special response
to collision it is unrealistic to assume that collisions are elastic,
or that the flagella adopt immediately to the new orientation which
changes during collisions.  Thus it
is reasonable to assume strong correlation between the cell's
orientation before collision and the cell's orientation after
collision. In addition, the orientation after the collision should be
biased according with the average direction of motion of the
surrounding bacteria, as they carry the liquid with them.
The important parameter is not the collision length $l_c$
but re-orientation time $\tau_r$. The re-orientation time is the time
it takes a bacterium to loose memory of its initial orientation, i.e.
the time span on which the final orientation has effectively no
correlation with the initial orientation. At low densities the
re-orientation time $\tau_r$ is equal to the tumbling time $\tau_T$.
As the density rises and the collisions become more frequent, $\tau_r$
decrease.  $ \tau_r$ defines the densities above which the constant
diffusion coefficient $D_b\equiv v^2\,\tau _T$ is not a good
approximation. It is quite possible that these densities are high
enough so as to make the velocity and even the type of motion
dependent on bacterial density, making relation (\ref
{eq:hige_density_diffusion:0}) irrelevant. In any case, high cellular
densities does mean an effective decrease in the diffusion coefficient
related to the bacterial movement.

When swimming in an unstirred liquid, very low cellular densities also
effect the movement. The bacteria secrete various materials into the
media and some of them, e.g. enzymes and other polymers, change
significantly
the physical properties of the liquid making it more suitable for
bacterial swimming. The secretion of these materials depend on
cellular density, thus at not-too-high densities the speed of swimming
rise with the cellular density. Hence the
diffusion coefficient related to the bacterial
movement should be a non-monotonic function of the bacterial density.
Moreover, the specific functional form might depend on the specific
bacterial strain.

In other conditions there is similar but more pronounced effect.
On semi-solid surface the bacteria cannot swim at all inside the agar
and they have to produce their own layer of liquid to swim in it.
A single \T bacterium on the agar surface cannot produce enough fluid
to swim in it, thus the bacteria cannot break out of the layer fluid
and the branches of a \T colony can be defined by this fluid.
Whenever bacteria enter the shallower parts of the layer, at the edge
of the branch, they become sluggish, indicating that the depth of the
layer effects the bacterial movement.
It can be argued (see section \ref{nonlinear}) that in such cases the
bacterial
speed is related to the bacterial density by a power law (at least in
low densities). Not only the diffusion coefficient related to the
bacterial movement is a non-monotonic function of the bacterial density
(as in a liquid agar), but it is also vanishes for extremely low
densities. 
In this case it is clear that the specific functional form depend on
the specific bacterial strain (\bsub-, for example, cannot move at all
under such conditions).

\subsection{Chemotaxis}

\label{sec:chemotaxis}

Chemotaxis means changes in the movement of the cell in response to a
gradient of certain chemical
fields~\cite{Adler69,BP77,Lacki81,Berg93}. The movement is biased
along the gradient either in the gradient direction or in the opposite
direction.
Bacteria are too short to estimate spatial gradients of the chemical
by simply comparing concentrations at different locations on their
membrane \cite{BP77} (but see Ref. \cite{Dusenbery98} for different
view). They deduce the spatial gradients by calculating
temporal derivatives along their path. It is known that {\it E.
coli}, for example, can compare successive measurements over a time
interval of $3$ seconds. The actual chemotaxis in swimming bacteria is
implemented by decreasing the tumbling frequency as cells swim up the
gradient of the attractant or down the gradient of repellent. Thus the
straight runs are important for gradient perception and the tumbling
timing is important for the response to this gradient.

Usually chemotactic response means a
response to an externally produced field like in the case of
chemotaxis towards food.  However, the chemotactic response can be also to a field
produced directly or indirectly by the bacterial cells. We will refer
to this case as chemotactic signaling.

Chemotaxis towards high concentration of nutrients is a well studied
phenomenon in bacteria \cite{Adler69,Adler66}. When the center of a
soft
agar plate (0.35\% agar concentration) is inoculated with
cells capable of chemotaxis, distinct circular bands of bacterial cells become visible after
a few hours of incubation. In fact, these patterns were used as
semi-quantitative indicators of chemotactic response \cite{Adler66}.
Genetic experiments showed that the creation of each of those bands
depends solely on the chemotactic response to a single chemical in the
substrate (these chemicals are usually metabolizable, but even
cells that have lost the ability to metabolize a certain chemical
form bands, as long as they are attracted to it \cite{Adler69}).
Berg \etal \cite{BB72} showed that the bacteria realize chemotactic
response by modulating the periods between tumbling events -- they
decrease the probability of tumbling when moving in a preferred
direction along the chemical gradient. This makes a bias in the random
walk which result in a mean drift of the bacteria in the desired
direction, a drift that can be as large as $v/10$.

Bacteria sense the local concentration $C$ of a chemical via membrane
receptors binding the chemical's molecules
\cite{Adler69,Lacki81}. The cell measures the concentration by
calculating the relative number of occupied receptors
$\frac{N_o}{N_o+N_f}$, where $N_o$ and $N_f$ are the number of
occupied and free receptors respectively. For a given chemical $C$,
$N_o$ is determined by two characteristic times: the mean time of a
receptor occupation -- $\tau _o $, a constant determined by internal cellular
processes -- and the mean time lapse when the receptor is free ($\tau
_f$). Since $\tau _f$ is inversely proportional to the concentration
of the chemical (with the proportion coefficient determined by the
receptor affinity to the chemical), we get:
\begin{equation}
\frac{N_o}{N_f+N_o}=\frac{\tau _o}{\tau _f+\tau _o}=\frac C{K+C}~,
\label{eq:bioBG:receptor}
\end{equation}
Where $K\equiv \left( C\,\tau _f\right) /\tau _o$ is constant.
It is crucial to note that when estimating gradients of chemicals, the
cells actually measure changes in the receptors' occupancy
$N_o/(N_o+N_f)$ and not in the concentration itself.
Using {Eq.~(\ref{eq:bioBG:receptor})} we obtain:
\begin{equation}
\frac \partial {\partial x}\left( \frac{N_o}{N_o+N_f}\right) =\frac
K{(K+C)^2 }\frac{\partial C}{\partial x}.
\label{eq:bioBG:receptorLow}
\end{equation}
This means that the chemical gradient times a factor $K/(K+C)^2$ is
measured. This dependence in known as the ``receptor law''
\cite{Murray89}.
For very high concentration the chemotaxis response vanishes due to
saturation of the receptors. The chemotactic response also vanishes at
the opposite limit of small concentration, as the concentration
reception is masked by external and internal noises. This effect is
not included in the receptor law, which should be changed
accordingly.

The receptor law is needed to explain the the bands reported by Adler
\cite{Adler69,Adler66}.
It can be shown that linear chemotactic response to a nutrient
cannot produce such bands. A non-linear response like
the ``receptor law'' must be included for the bands to form. Moreover,
high concentration of the attractant represses both the strength of
the chemotactic response \cite{Adler69} and the velocity of the
expanding band \cite{WB89}. These observations are accounted for by
the ``receptor law'' for chemotactic response if one assumes that the
average gradient sensed by the cells is proportional to the
initial concentration of the chemical \cite{Adler69,WB89}.

The bacterial flux due to chemotaxis can be described by
\begin{equation}
\vec{J}_{chem}\equiv \zeta (\sigma)\chi (C)\nabla C
\end{equation}
where $\chi (C)\nabla C$ is the gradient sensed by the cell (with
$\chi (C)$ having the units of 1 over chemical's concentration) and
$\zeta (\sigma)$ is the bacterial response to the sensed gradient
(having the same units as a diffusion coefficient). $\chi (C)$ is
usually taken to be either constant or the ``receptor law''.

The function of the bacterial response $\zeta $ is positive for
attractive chemotaxis (movement towards high concentrations) and
negative for repulsive chemotaxis. If the movement is in liquid and at
low bacterial densities, then $\left| \zeta (\sigma)\right| \propto
\sigma D_b$. In a lubrication fluid which effect the bacterial
movement, the chemotaxis is effected in the same way the diffusion is;
$\left| \zeta (\sigma ,l)\right| \propto \sigma D_b(\sigma ,l)$.

In the case of high bacterial densities, collisions between
bacteria can disrupt both the perception of chemical gradient and the
bacterial response. As the collisions prevent the bacteria from moving
on a straight line between tumbling events, the effective response to
chemotaxis is reduced.

\subsection{Food Consumption, Reproduction and Starvation}

\label{sec:eating}

The {${\cal T}$} bacteria, like most bacteria, reproduce by fission of the
cell into two daughter cells which are practically identical to the mother
cell. The crucial step in the cell division is the replication of the
genetic material and its sharing between the daughter cells. Haste
replication of DNA might lead to many errors -- most organisms limit
the rate of replication to about 1000 bases per second.  Thus the
reproduction must take at least minimal reproduction time $\tau _R$.
This reproduction time $\tau _R$ is about $25min$ in Bacilli.

For reproduction, as well as for movement and other metabolic
processes, bacteria and all other organisms need influx of energy. Any
organism which does not get its energy directly from sunlight (by
photo-synthesis) needs an external supply of food. In the patterning
experiments the bacteria eat nutrient from the agar. As long as there
is enough nutrient and no significant amount of toxic materials, food
is consumed (for cell replication and internal processes) at maximal
rate $\Omega _c$. To estimate $\Omega _c$ we assume that a bacterium
needs to consume an amount of food $ C_R$ of about $3\times
10^{-12}g$.  It is 3 times its weight -- one quanta for doubling body
mass, one quanta used for movement and all other metabolic processes
during the reproduction time $ \tau _R$, and one quanta is for the
reduced entropy of making organized cell out of food.  Hence $\Omega
_c$ is about 2$fg/sec$ (1$fg=10^{-15}$gram).

If nutrient is deficient for a long enough period of time, the
{${\cal T}$} cells may enter a special stationary state -- a state of
a spore -- which enables them to survive much longer without food.
The bacterial cells employ very complex mechanisms tailored for the
process of sporulation.
They stop normal activity -- like movement -- and use all their
internal reserves to metamorphose from an active volatile cell to a
sedentary durable 'seed'. While the spores themselves do not emit any
chemicals (as they have no metabolism), the pre-spores (sporulating
cells, see Fig. \ref{fig:spores}) do not move and emit a very wide range of waste materials,
some of which unique
to the sporulating cell. These emitted chemicals might be used by
other cells as a signal carrying information about the conditions at
the location of the pre-spores. Ben-Jacob {\it et al.}
\cite{BSTCCV94a,BSTCCV94b,CCB96}
suggested that such materials are repelling the bacteria ('repulsive
chemotactic signaling') as if they escape a dangerous location.

When bacteria are grown in a petri dish, nutrients are usually
provided by adding peptone, a mixture including all the amino acids
and sugars as source of carbon. Bacteria which are not
defective in synthesis of any amino acid can grow also on a
minimal agar in which a single source of carbon and no amino acids
are provided.
Such growth might seem to be easier to model as the growth is limited
by the diffusion of a single chemical. However, during growth on
minimal agar there is usually a higher rate of waste products
accumulation, introducing other complications into the model.
Moreover many of our strains are auxotrophic i.e. defective in
synthesis of some amino acids and need an external supply of it.
Providing the bacteria with these amino acids and only a  single
carbon source might pose us the question as to what is the limiting
factor in the growth of the bacteria.  For all those reasons we prefer
to use peptone as nutrient source.

We said that if there is ample supply of food, bacteria reproduce in a
maximal rate of one division in $\tau_R$. If the available amount of
food is limited, bacteria consume the maximum amount of food they can.
In the limit of low bacterial density, the available amount of food
over the tumbling time $\tau _T$ is the food contained in the area
$\tau _T\sqrt{D_bD_n}$, where $D_b$ and
$D_n$ are the diffusion coefficients of the bacteria and the food,
respectively. Hence the rate of food consumption is given by
$n\sqrt{D_bD_n}$ (weather $D_b$ is constant or not).

In a continuous model, reproduction of bacteria translate to a growth
term of the bacterial density which is $\sigma $ times the eating rate
per bacteria. In the limit of high nutrient it is
$\sigma /\tau_R$, and in the limit of low nutrient it
is proportional to $n\sigma $. This brings to mind Michaelis-Menten
law \cite{Murray89} of $\frac K{1+\gamma n}n\sigma $ with $K,\gamma $
constants. Many authors take only the low nutrient limit of this
expression, $K n\sigma$, although it is not biologically established
that the bacteria in the experiments are limited by the availability
of food and not by their maximal consumption rate.

\section{Reaction-Diffusion Models}
\label{sec:reaction}
In this section we deal with continuous, reaction-diffusion models for
bacterial growth. The models under study are due to Fisher and
Kolmogorov \cite{Fisher37,KPP37}, Kessler and Levine \cite{KL98},
Kitsunezaki \cite{Kitsunezaki97}, Kawasaki \etal  \cite{KMMUS97} and
Mimura \etal \cite{Mimura97,MWIRMSM98}.
The models are all two-dimensional (2D),
with  $b({\bf x},t)$ denoting the density of bacteria projected on a
2D plane and  $n({\bf x},t)$ is the 2D nutrient density. The equations for the various models  will be written in dimensionless units, and the
reader is referred to the Appendix for a discussion about the relations
with real units. 

In general, the rate of change of the bacteria density can be
described by\cite{Murray89}:
\begin{equation}
\label{murray-eqn}
\partderiv{b}{t} = movement + ``birth'' - ``death''
\end{equation}
As discussed in Sec.\ \ref{sec:bio-background}, the movement of bacteria
consists of various possible mechanisms, of which we will concentrate
on swimming, so that the motion is described as diffusion (either
linear or non-linear). The ``birth'' term in Eqn.\ \ref{murray-eqn}
corresponds to bacterial
reproduction, which depends on the supply of nutrients. The 
``death'' term represents the transition of bacteria into a non moving
state.

\subsection{The Fisher-Kolmogorove equation}
\label{sec:fisher}
Mathematically, the above description is usually written as a {\em reaction diffusion
  equation}, for which the canonic example is the Fisher-Kolmogorove
equation \cite{Fisher37,KPP37} (without a death term):
\begin{equation}
\label{fisher-eqn}
\partderiv{b}{t} = D_b \nabla^2 b + b(1-b)
\end{equation}

This equation was originally presented to describe the
spread of mutants in a population. We will use it here as a
starting point for our discussion of colonial development. In this context, $D_b$ is the
diffusion coefficient describing the bacterial movement, and the
reaction term $f(b)=b(1-b)$) describes both the growth and ``death''
of bacteria.
 The function $f(b)$ is
depicted in Fig.\ \ref{fisher-fig}. 
Eqn.\ \ref{fisher-eqn} has two homogeneous solutions, a stable solution $b=1$
and an unstable solution $b=0$. These solutions
correspond to the two extrema of the potential  $\Phi = - \int{f(b)db}$,
  which can be thought of as a Landau-Ginzburg free energy density (see
  Fig.\ \ref{fisher-fig}). Thus, we can study
the propagation of the stable state (inside the colony) into the
unstable one (outside the colony). It can be shown \cite{AW75,BBDKL85,vanSaa88,vanSaa89,cross,KNS98} that
in 1D there is a unique selected velocity
of the front, chosen according to the properties
of the fixed point far ahead at infinity.

In two dimensions, the propagation is in the form of a compact (as
opposed to branching) growth. That is, there is no Mullins-Sekerka
instability \cite{BTSA94,MS64,MWM95,BSTA95}.
In the case of such instability, a small bump in a flat interface will
have a higher velocity than the rest of the front and will therefore over-grow.
Here, however, this will not happen, because the front velocity is determined by the  
fixed point at infinity, rather than by local properties of
the front.

The Fisher-Kolmogorove model is an appropriate description of the
growth when bacteria are grown
under nutrient-rich conditions, in which case the growth dynamics is
not limited by the supply of food \cite{WKNM94}. Here we are interested in the opposite case, where nutrient
supply is limited. 
For a more realistic description of colonial development on a
nutrient-poor surface, we must take
into account the interaction of bacteria with the nutrient field
$n({\bf x},t)$. In the simplest case, this is described by the {\em
  Diffusive Fisher-Kolmogorove equation}\cite{MWM95,BSTA95}:
\begin{eqnarray}
  \label{diffusive-fisher-eqn}
  \lefteqn{ \partderiv{b}{t} = D_b \nabla ^2  b + f(b,n) }\\
  \lefteqn{ \partderiv{n}{t} = D_n \nabla ^2 n - \eta f(b,n)}
\end{eqnarray}
where $\eta>0$ is the conversion ratio of food into bacteria (3
picogram per bacteria, see Sec.\ \ref{sec:bio-background}). In this model the
food consumed by the bacteria is reduced from the food field and
converted into bacteria. The shape of a one-dimensional front obtained for $f(b,n)=bn$ is
depicted in Fig.\ \ref{fisher-front-fig}.
It turns out that this model (as the original
Fisher-Kolmogorove model) has a selected front velocity, determined by
the conditions at infinity, and therefore
 does not exhibit diffusive (Mullins-Sekerka) instability, and a two
dimensional growth is compact rather than branched.
An additional way of understanding why compact growth is obtained even
under poor nutrient levels is to note that the model, as it is, does
not impose a minimal bacteria density in the colony, so that the
density can be adjusted according to the initial food level. In the real
biological system, however, some minimal density is required in order
to create the lubricating fluid (needed for movement), and so compact
growth usually is not possible.

For further understanding the requirements for a branching pattern, let us recall the case of
solidification from an under-cooled melt, which exhibits
branching patterns \cite{BG90,BenJacob93}.
Our bacterial density $b$ corresponds to the order parameter in the
phase-field model description of solidification \cite{KSB94,Kupferman95}, whereas
the diffusion of food is analogous to the diffusion of heat away from
the solid-liquid interface. In the case of solidification,
the Landau-Ginzburg free energy is a tilted double-well (see Fig.\ \ref{meta-fig}).
The meta-stable state corresponds to the liquid phase. In the
diffusive Fisher-Kolmogorove case, the analog of the liquid state,
i.e. the $b=0$ state, is unstable. Thus, according to the
solidification case, if we modify the model and turn the $b=0$ state
into a meta-stable one, this can lead to branching growth.

\subsection{A cutoff in the reaction term} 
\label{sec:cutoff}
In the case of bacteria, there is a feature of the system that
might have a similar effect to the meta-stability in
solidification.
This is the discreteness of bacteria, for which the continuous
description is not always valid. Kessler and Levine \cite{KL98}
argue that when describing a discrete system using continuous models,
a cutoff near the fixed point must be imposed, i.e. the
reaction term must be set to zero when the (bacterial-) density is
below some threshold. They have shown that inclusion of such a cutoff 
leads to a Mullins-Sekerka instability, and branching patterns may
appear when a death term is also included, as explained below.

To show this, we consider the diffusive Fisher equations with a
cutoff:
\begin{eqnarray}
\label{kl-eqn}
\lefteqn{ \partderiv{b}{t} = D_b \nabla ^2  b + b n \Theta(b-\beta) }\\
\lefteqn{ \partderiv{n}{t} = \nabla ^2 n - n b \Theta(b- \beta)}
\end{eqnarray}   
where $\beta$ is the threshold density for growth, and $\Theta$ is
the Heaviside step function.
The food consumption term is of the form $f(n,b) = n b$,
which is the widely used low-nutrient
approximation for the Michaelis-Menten law (Sec.\ \ref{sec:eating}).
The value of the threshold for bacterial growth, $\beta$, is taken to
be one bacterium per 1-10 $\mu m ^2$. Note that this corresponds to the
case where the distance between bacteria is of the order of the length
between tumbles.
Fig.\ \ref{kl-fig} depicts the reaction term and the Landau-Ginzburg free
energy for this model. As can be seen in Fig.\ \ref{kl-pattern-fig},
an instability of the front indeed appears, and the compact growth pattern has
a surface broken by ``fjords''. However, the Mullins-Sekerka instability is not
sufficient to produce branches. The emerging dips soon
``heal'', so that branches are not formed.  
One way to obtain branching growth is to add a  ``death'' term to the
model, thus:
\begin{eqnarray}
\label{kl-death-eqn}
\lefteqn{ \partderiv{b}{t} = D_b \nabla ^2  b + b n \Theta(b-\beta) - \mu b}\\
\lefteqn{ \partderiv{n}{t} = \nabla ^2 n - n b \Theta(b- \beta)}\\
\lefteqn{ \partderiv{s}{t} =  \mu b}
\end{eqnarray}
where $\mu$ is the rate of bacterial differentiation into non-moving state, and $s({\bf x},t)$
is the density of ``frozen'' bacteria.
This modified model exhibits distinct branching patterns, as seen on
Fig.\ \ref{kl-death-fig}.
The explanation for this effect lies in the fact that now, with a
death term present, bacteria left behind the propagating front become
non-motile (``dead''). They are unable to move and close the
``fjords'', thus allowing real branches to form.

\subsection{Reaction-Diffusion with Lubrication}
\label{sec:lubrication}

We have so far ignored the effect of the lubricating 
field on the motion of the bacteria. We present here a new model which 
 incorporates an additional field that
describes the lubricating fluid. The field, denoted as $l$, is the local height of the lubrication fluid
on the agar surface. Its dynamics is governed by a reaction diffusion equation. 
The bacterial diffusion is coupled to this field.

The dynamics of the field is given by:
\begin{equation}
\partderiv{l}{t}=-\nabla  \vec{J_l}+\Gamma bn(l_{max}-l) - \lambda l
\end{equation}
where $\vec{J_l}$ is the fluid flux 
(to be discussed), $\Gamma$ is the production rate and $\lambda$ is the 
absorption rate of the fluid by the agar.

We assume that the fluid production depends on the
bacterial density.
As the production of lubrication probably demands substantial energy, 
it should also depend on the nutrients level. 
We assume that the absorption of fluid into the
agar depends on the local amount of fluid (i.e. the height of the fluid layer).
In this model the production depends linearly on
the concentrations of both the bacteria and the nutrients.
The production term cannot become negative as the lubrication height
cannot exceed $l_{max}$.

The lubrication fluid flows by diffusion and by convection caused by bacterial motion.
A simple description of the convection is that each bacterium drags along its movement
the fluid surrounding it. 
\begin{equation}
\vec{J_l}=-D_l l^\eta \nabla l + j\vec{J_b}
\end{equation}
where $D_l$ is the lubrication diffusion coefficient, $\vec{J_b}$ is the 
bacterial flux and $j$ is the amount of fluid dragged by each bacterium.
The diffusion term of the fluid depends on the height of the fluid to the power $\eta$.
The nonlinearity causes the fluid
to have a sharp boundary at its front, as is observed in the experiments
of bacterial colonies development.

We now turn to the effect of the lubrication on bacterial diffusion.
An increase in the amount of lubrication decreases the friction 
between the bacteria and the agar surface. 
The term 'friction' is used here in a very loose manner
to represent the total effect of any force or process that slows down
the bacteria. It might include, for example, the drag which acts on a
body moving in shallow layer of viscous fluid. It might include the
probability that a flagellum will adhere or get tangled with the
polymers of the agar.
As the bacterial motion is over-damped, the local speed of the bacteria
(or the mean step length, when neglecting collisions between
bacteria) is proportional to the self-generated propulsion force
divided by the friction.
It can be shown that variation of the step length leads to variation of
the diffusion coefficient, with the diffusion coefficient proportional
to the step length to the power of two.
We  assume that the friction is inversely related to the local lubrication 
height through some power law: $friction\sim l^{\gamma} \mbox{ and } \gamma < 0$.
Following our arguments the bacterial flux is:
\begin{equation}
\vec{J_b}=-D_b l^{-2\gamma} \nabla b
\end{equation}
For the complete model we took simple bacterial growth and death terms. The model is:
\begin{eqnarray} 
\partderiv{b}{t}&=&D_b\nabla (l^{-2\gamma} \nabla b)+ bn-\mu b\\
\partderiv{n}{t}&=&\nabla^2n-bn \nonumber\\
\partderiv{l}{t}&=&\nabla (D_l l^\eta \nabla l+jD_b l^{-2\gamma} \nabla b)
+\Gamma bn(l_{max}-l) - \lambda l \nonumber\\
\partderiv{s}{t}&=&\mu b \nonumber\\
\end{eqnarray}
For the initial condition, we set:
\begin{eqnarray} 
  n({\bf x},t)&=&n_0 \\
  b({\bf x},t)&=&b_0({\bf x}) \nonumber
\end{eqnarray}
where $n_0$ is the initial concentration of nutrients and $b_0({\bf x})$
is the initial bacterial distribution. In the simulations, $b_0({\bf x})$ is  localized at the center.

Preliminary results show that the model can reproduce branching
patterns (Figure \ref{yony-fig}).
At low values of the absorption rate
the model exhibits dense fingers. 
At higher values of the absorption rate 
the model exhibits finer branches. We obtain finer branches also if we change other parameters
that effectively decrease the amount of lubrication. 
We can relate these conditions to high agar concentration.
In this model, as in the non-linear diffusion model described below,
the bacterial field and the lubricating field have a front ``wall'' with compact support
(Figure \ref{yony-front}).

\subsection{Non-linear diffusion}
\label{nonlinear}
It is possible to introduce a
simplified model, where the fluid field is not included, and is replaced
 by a density-dependent
diffusion coefficient for the bacteria $D_b \sim b^k$ \cite{Cohen97,BLC98}.
For this purpose, a few assumptions 
 are needed about the dynamics at low
bacterial and lubrication density:
\begin{itemize}
\item The production of lubricant is proportional to the bacterial
  density to the power $\alpha >0$.
\item There is a sink in the equation for the time evolution of the
  lubrication field, e.g. absorption of the lubricant into the agar.
  This sink is proportional to the lubrication density to the power
  $\beta >0$.
\item Over the bacterial length scale, the two processes above are
  much faster than the diffusion process, so the lubrication density
  is proportional to the bacterial density to the power of $\beta /
  \alpha$.
\item The friction is proportional to the lubrication density to the
  power $\gamma < 0$.
\end{itemize}
Given the above assumptions, the lubrication field can be removed from
the dynamics and be replaced by a density dependent diffusion
coefficient. This coefficient is proportional to the bacterial density
to the power $k \equiv -2 \gamma \beta / \alpha > 0$

A model of this type is offered by Kitsunezaki\cite{Kitsunezaki97}:

 \begin{eqnarray}
\label{kitsunezaki-eqn}
\lefteqn{ \partderiv{b}{t} = \nabla (D_0 b^k \nabla
  b) + n b - \mu b}\\
\lefteqn{ \partderiv{n}{t} = \nabla ^2 n - b n}\\
\lefteqn{ \partderiv{s}{t} = \mu b}
\end{eqnarray}

For $k>0$ the 1D model gives rise to a front ``wall'',
with compact support (i.e. $b=0$ outside a finite domain, see Fig.\ \ref{kits-k1-fig}). For
$k>1$ this wall has an infinite slope (Fig.\ \ref{kitz-k2-fig}).
The propagation velocity in this case is determined by the condition
at the front, not at infinity \cite{Cohen97,Newman83}. We therefore expect a Mullins-Sekerka
instability in 2D (as is claimed in \cite{Kitsunezaki97}). Indeed, the
model exhibits branching patterns for suitable parameter values and
initial conditions
(Fig. \ref{kitz-pattern-fig}).
Note, that the compact support exhibited by this model (that is, the
abrupt disappearance of bacterial presence outside the colony
boundary) is much more in
accordance with experimental observations than the long ``tail'' of
bacterial density appearing in the Fisher-Kolmogorove case (recall
Fig.\ \ref{fisher-front-fig}).
 
Another state-dependent diffusion coefficient was proposed and studied by Kawasaki
\etal \cite{KMMUS97}, which took $D_b \sim n b$. They justify this form
by the observation that bacteria
are active mostly at the edge of the colony -- the only area where
there is high amount of bacteria {\em and} food. 
Their model, too, exhibits
branching shapes (see Fig.\ \ref{kawa-pattern-fig}).
This is due to the $b$ dependence of the diffusion coefficient,
which leads to front instability, just as in the Kitsunezaki model.
The fact that $D_b$ also depends on $n$ prevents bacteria
inside the colony from moving -- and closing the dips created by the
instability. In this way, branches are created
without a need for a death term.
A similar mechanism of ``food'' dependent diffusion
coefficient was used by Tu \etal \cite{TLR93}, who describe a mean-field
model for DLA. Their model, too, does not include a death term yet produces branching patterns.

\subsection{A meta-stable reaction term}
\label{metastable}

As mentioned earlier, meta-stability of the growth term can lead to
branching patterns. Mimura \etal \cite{Mimura97,MWIRMSM98} have
studied the following model, for which $b=0$ is a meta-stable fixed point:
\begin{eqnarray}
\label{mimura-eqn}
\lefteqn{ \partderiv{b}{t} = D_b \nabla ^2  b + \epsilon b n  -
  \frac{\mu b}{(b+1)(n+1)}}\\
\lefteqn{ \partderiv{n}{t} = \nabla ^2 n - n b}\\
\lefteqn{ \partderiv{s}{t} =  \frac{\mu b}{(b+1)(n+1)}}
\end{eqnarray}
(see the Appendix for relations to real units).
In this model, $D_b$ does not depend on bacterial density, and
it's value is said to vary with agar concentration. The key
feature of this model is that the total bacteria growth term
(multiplication minus inactivation), depicted in Fig.\ \ref{mimura-fig}, gives a meta-stable state at b=0. This means that in
order to initiate bacterial growth, a threshold value of
$b^{\ast}=\frac{\mu}{\epsilon n (n+1)} - 1$ must be reached (this
value corresponds to bacterial density of the order of 0.1
bacteria/$\mu m ^2$).
This feature contradicts the observations, that even a small number of
bacteria, inoculated on a substrate, will multiply and later begin to
move.
This does not imply that the model is incorrect.
A possible interpretation is that the field $b$
actually describes a combination of lubricant + bacteria. In this
case, however,
as we have explained before, we would expect the model to exhibit a non-linear diffusion.
Hence we believe that this model might provide a better description of
the bacteria if the diffusion term is replaced with a non-linear
diffusion term.
The model as it is exhibits various branching patterns, patterns of concentric
rings and compact growth (Fig.\ \ref{mimura-patterns-fig1}).

Mimura \etal argue that the model captures  the experimental morphology diagram they observed.
This is a very crucial point. If indeed the above claim is correct,
it implies that the observed patterns can be reproduced with no need
for additional biological features. However, using the discrete communicating walkers
model \cite{BSTCCV94a}, Ben-Jacob \etal have concluded hat the additional
features of chemotactic response have to be included. So, in order to
check this point we have performed more detailed comparison between the
Mimura \etal model and the experimental observations. 

First, we consider the DLA-like growth. In this case, the bacteria do
not move on the agar surface, and the growth is indeed very similar to
the DLA algorithm, as was proposed by Matsushita \etal \cite{WitSan81,MF90}. It is now understood that in a mean-field DLA model
the particle density (density of bacteria in the present case) can not be
described by a diffusion term. Instead, it has to be described by a
diffusion multiplied by the nutrients field \cite{TLR93}, which differs from the
linear diffusion in the model of Mimura \etal.
Indeed, close inspection of the fractal pattern created by the model
reveals that it differs from the observed DLA-like patterns.

Another test of the model is the predicted pattern of concentric
rings. It has already been pointed out by Rafols \cite{Rafols98} that
the model's pattern differs from the observed one. In the experiment,
branching growth slows down. The branches become wider and growth
stops. Then, after bacterial differentiation, a new cycle of branches
growth starts with thin branches emitted from the stationary wide
branches \cite{Rafols98}.
This description differs from the model patterning shown in
Fig. \ref{mimura-patterns-fig1}.

In Fig. \ref{mimura-patterns-fig2} we exhibit results of numerical
simulations for various levels of peptone and for agar concentration
for which concentric rings are observed. The sequence of patterns from
DLA-like at low peptone to concentric rings at high levels of peptone
differ from the similar sequence of observed patterns presented in
Ref. \cite{Rafols98}.

We have also tested the change in patterns as we vary the agar
concentration
(see Fig. \ref{mimura-patterns-fig3}). When we plot the growth velocity as a
function of the agar concentration, it does not show a jump in the
velocity or its slope. In other words, the model seems to exhibit a
simple crossover between the patterns rather then a morphology
transition as the observations seem to indicate.

The above results lead us to conclude that the Mimura \etal model does
capture some of the observed branching patterns, yet the complete
description of the observations requires additional features to be
included. Specifically, we propose to include nonlinear diffusion. We
also believe that chemotactic response does play an important role for
poor growth conditions.
To further test this conclusion, we present in the next
section a study of patterns produced by the reaction-diffusion models
when chemotactic responses are included.  

\section{Incorporation of Chemotactic Signaling in the Continuous
  Models}
\label{sec:incorporation}

So far we have seen several models for the branching colonies, each
with its own mechanism of diffusive instability, which produces
patterns resembling the observed ones. Is there a way to
distinguish between the models so as to find out what are the
biological features underlying the diffusive instability?
We rely on results from an atomistic model -- the
Communicating Walkers model \cite{BSTCCV94a,BSCTCV95,CCB96} -- for an
indication of what are the biological features relevant to the
different morphologies.

\subsection{Chemotactic-based branching growth: Insights from the Communicating
Walkers model}

 Ben-Jacob \etal argued
that for the colonial adaptive self-organization the \Tm employs
several
kinds of chemotactic responses.
Usually chemotactic response means a bias of the movement in response
to a gradient of an externally produced field like in the case of food
chemotaxis.
However, it could also be a response to a field produced directly or
indirectly by the bacterial cells -- chemotactic signaling.

Three kinds of chemotactic responses are suggested to be employed by
the \T morphotype, each dominant in different regime of the morphology
diagram.
One response is the food chemotaxis we have mentioned earlier.
According to the "receptor law", it is expected to be dominant for 
some range of nutrient levels (the corresponding levels of peptone are
determined by the constant $K$). The two other kinds of chemotactic
responses are signaling chemotaxis. One is long-range repulsive
chemotaxis where the
chemical is secreted by starved bacteria at the inner parts of the
colony. The second signal is a short-range attractive chemotaxis
where the chemical is secreted by bacteria at the colony's front,
bacteria which are immersed in toxic waste products.
The length scale of each signal is determined by the diffusion
constant of the chemical agent and the rate of its spontaneous
decomposition. 

{\em Amplification of diffusive Instability Due to Nutrients Chemotaxis:}
In non-living systems, the more ramified patterns
(lower fractal dimension) are observed for lower growth velocity.
Based on growth velocity as function of peptone level and based on
growth dynamics, Ben-Jacob \etal concluded that in the case of
bacterial colonies there is a need for mechanism that can both
increase the growth velocity and maintain, or even decrease, the
fractal dimension.
Ben-Jacob \etal suggested food chemotaxis to be the required
mechanism.
It provides an outward drift to the cellular movements; thus, it
should increase the rate of envelope propagation. At the same
time, being a response to an external field it should also amplify the
basic diffusion instability of the nutrients field. Hence, it can
support faster growth velocity together with a ramified pattern of low
fractal dimension.

The Communicating Walkers model was used to test the above hypothesis.
The simulations showed that as expected, the inclusion of food
chemotaxis in the model led to a considerable increase of the growth
velocity without significant change in the fractal dimension of the
pattern.

{\em Repulsive Chemotactic Signaling: }
We focus now on the formation of the fine radial branching patterns at
low peptone levels. From the study of non-living systems it is
known that, in the same manner that an external diffusion field leads
to the diffusion instability, an internal diffusion field will
stabilize the growth.
It is natural to assume that such a field is produced by some sort of
chemotactic agent. To regulate the organization of the branches, it
must be a long-range signal. To result in radial branches it must be a
repulsive chemical produced by cells at the inner parts of the
colony. The must probable candidates are the bacteria 
entering a pre-spore stage due to depletion of nutrient. 
This proposal has also been verified by simulations of the 
 Communicating
Walkers model. In the presence of repulsive chemotaxis
the patterns become much denser with a smooth
circular envelope, while the branches are thinner and radially
oriented.

\subsection{Results for the continuous models}
\label{sec:results}
We incorporate the effect of chemotaxis in the continuous models by
introducing a {\em chemotactic flux} $\vec{J}_{chem}$, which is written (for
the case of a chemorepellent and a linear diffusion) as
\cite{Murray89}:

\be
{\vec{J}_{chem} = - b \chi (r) \nabla r}
\label{j_chem_linear}
\ee
where $r({\bf x},t)$ is the concentration of chemorepellent and 
$\chi(r)$ is the chemotactic sensitivity to the repellent.
In the case of a chemoattractant, e.g. a nutrient, the expression for
the flux will have an opposite sign.

Recall that in the case of the ``receptor law'', the sensitivity $\chi(r)$ takes the form:

\be
{\chi(r) = \frac{\chi_0 K}{(K+r)^2}}
\ee

Thus, we obtain the {\em reaction diffusion chemotaxis equation}:

\be
{\partderiv{b}{t} = \ -\nabla  (-D \nabla b - b \chi(r) \nabla r)
  + f(b)}
\ee

In addition, one has to write an equation describing the diffusion and
the production and decomposition of the chemorepellent.
This is written as follows:
\be
{\dot{r} = D_r \nabla^2 r + \Gamma_r s - \Omega_r b r - \lambda_r r}
\label{r-eqn}
\ee
where $D_r$ is the diffusion coefficient of the chemorepellent,
$\Gamma_r$ is the emission rate of repellent by pre-spores, 
$\Omega_r$ is the decomposition rate of the repellent
by active bacteria,
and $\lambda_r$ is the rate of self decomposition of the
repellent.

We have tested the effect of food chemotaxis and
repulsive chemotaxis in several of the aforementioned reaction-diffusion
models,
and present here results for the Kitsunezaki nonlinear
diffusion model and the Mimura \etal meta-stable model.

When treating the Non-linear diffusion model, we must modify the
expression for the chemotactic flux, in order to incorporate the 
density dependence of the bacterial movement. Thus, similarly to the
diffusion coefficient, which is written as $D_b = D_0 b^k$, 
we modify the chemotactic flux to become:
\be
{\vec{J}_{chem} = \zeta(b) \chi (r) \nabla r}
\label{j_chem}
\ee
where $\zeta(b) = b \cdot b^k = b^{k+1}$ is the bacterial response to
the chemotactic agent. For the linear diffusion case, $\zeta(b)$
degenerates again to $b$.

Fig.\ \ref{kits-food-fig} depicts a pattern developed by the
Kitsunezaki model when food chemotaxis is included. All of the parameters are the
same as in Fig.\ \ref{kitz-pattern-fig} (no chemotaxis). 
Although  the patterns are very similar, the growth velocity
when food chemotaxis is included is about twice the velocity in the
absence
of chemotactic response. In other words, the
velocity is doubled with no change in the fractal dimension.

The effect of repulsive chemotactic signaling is demonstrated in
Fig.\ \ref{kits-rep-fig} -- again with otherwise identical parameters
to those in Fig.\ \ref{kitz-pattern-fig}. 
It can be seen that the previously fractal-like shape has turned
into a radial  branching pattern with a circular envelope.

Thus, for the two types of chemotactic response -- food and repulsive,
the results we observe are similar to those  obtained for the
communicating walkers model. This agreement is not surprising,
as
both models capture the important feature of a
lubrication fluid. Recall also that the nonlinear diffusion
model is the only one which exhibits a sharp front in 1D.

The effect of chemotactic response in the
Mimura \etal model is presented in
Figs.~\ref{mimura-food-fig} and \ref{mimura-rep-fig}.
The chemotactic response was added to a previously DLA-like colony
(Fig.\ \ref{mimura-patterns-fig1} a). The
addition of food chemotaxis turns the colony into a densely branched
one, with branches much thicker than before. The repulsive chemotaxis
makes the branches radially oriented, but they become thicker
than before. Thus, the effect of chemotactic responses in this model
differs from the one obtained for both the walker model and the
Kitsunezaki model. We believe this to stem from the fact that the
model does not 
include nonlinear diffusion to capture the effect of lubrication.

\section{Discussion}
\label{sec:discussion}

We first briefly reviewed experimental observations of branching
patterns in various bacterial strains, under a range of growth
conditions.
Both the colonial patterns and the optical microscope observations of
the bacterial dynamics were presented. We have also included a brief
summary of the known key biological features required, as we think,
for successful modeling of the growth. 

Our goal in this manuscript was to test reaction-diffusion
models. To this end we surveyed the reaction-diffusion models for
branching growth that we are familiar with. Mathematical analysis
reveals that a number of different features can lead to instability of
a propagating front: A density-dependent (i.e. nonlinear) diffusion coefficient;  a lower
cutoff in the growth term, and a meta-stable growth term.
For this instability to create pronounced branches, a death term must
be added to the growth. Such a term prevents bacteria inside the colony from moving and
 ``healing'' the dips on the surface. Making the diffusion term
nutrient dependent can lead to a similar effect.
The experimental observations provide a clear indication that the
bacteria turn immobile, so that a ``death'' term indeed has to be
included in the models.

The fact that different mathematical features can lead to similar (to
the eye) branching patterns emphasizes how cautious we have to be in
modeling the colonial development. True, it might be that 
different bacterial strains develop branching patterns by the
employment of different biological features (each might correspond to
one of the mathematical features). Yet, when we consider a specific
bacterial strain, comparing the experimentally observed and model's
patterns is not sufficient to tell us if indeed the right biological
features are included in the model.

What, then, is the right way to tackle the problem? The integrative
way:
One must combine the mathematical knowledge (in this case, what mechanisms
lead to instability and branching) with an attempt to model the
(assumed) generic biological features, and comprehensive
comparison with experimental results.
As an example, the (mathematical) mechanism of non-linear diffusion
leads to a branching colony, with compact support. We must then ask
ourselves whether the movement of bacteria could be described in such
a way.
The answer is positive. As we have explained earlier, the lubrication
fluid can be modeled by a nonlinear diffusion of the bacteria.

For more critique test of the models, additional aspects of the growth
(such as functional dependence of the colonial growth velocity on
growth conditions, branches size and width distributions etc.) have to
be compared with the model predictions.
One should also compare the theory with more involved experimental
tests, such as the effect of imposed anisotropy, competition between
neighboring colonies, and expression of mutants (emergence of
sectors) in expanding colonies. 

Our conclusion from the study of bacterial branching growth is that
the minimal features of diffusion, food consumption,
reproduction and
inactivation are not sufficient to explain the complete picture of the
observed phenomena. We believe that additional mechanisms must be introduced,
and propose chemotactic signaling as plausible one.

This work has dealt with continuous models. Such models are preferable to discrete ones.
Each has its advantages and disadvantages. The discrete walkers model, for
example, enables us to include the valuable feature of internal degrees
of freedom, but is computationally limited in the number of walkers
that can be simulated, and thus its scaling to the real problem is
somewhat problematic.
The best strategy is to employ in parallel both the reaction-diffusion
and the walkers approach.

To conclude, despite the difficulties and possible pitfalls, we hope
to attract researchers to this emerging new endeavor. After all, a
significant progress has been made towards working out the cybernetic
processes (communication, regulation and control) during colonial
development. Yet, many challenging puzzles are waiting to be solved
and tantalizing phenomena are waiting to be discovered.

\section*{Acknowledgments}
We have benefited from many discussions on the presented studies with
H.\ Levine. We thank M.\ Matsushita for discussions about the Mimura
et \etal model. We are greatfull to I.\ Rafols for sending us a copy of
his master thesis which we found most helpful. IG wishes to thank R.\
Segev for fruitful discussions.
Identifications of the
\Tm and genetic studies are carried in collaboration with the group of
D.\ Gutnick. Presented studies are supported in part by a grant from
the Israeli Academy of Sciences grant no.\ 593/95 and by the
Israeli-US Binational Science Foundation BSF grant no. 00410-95.

\section*{Appendix: From Real Units to Dimensionless Models}
\label{sec:scaling}

Our goal is to relate the dimensionless equations with biophysical values of the parameters.
The procedure is to set the dimensional units to be the natural scales.
\subsection{Deriving the equations}
The diffusive Fisher-Kolmogorove equations in dimensional form 
(section \ref{sec:fisher}) are:
\begin{eqnarray} 
\label{start}
\partderiv{b}{t}&=&D_b\nabla^2b+E_bnb \\
\partderiv{n}{t}&=&D_n\nabla^2n-E_nnb \nonumber
\end{eqnarray}
$b$ and $n$ are the two dimensional concentration of bacteria and of nutrition 
respectively. 
The field $b$ is measured in units of number of bacteria per area in $cm^2$. 
The field $n$ is measured in units of grams per area in $cm^2$.
Experimentally nutrients are usually measured in $gram/liter$.
Note that $1g/l$ corresponds to $0.3 mg/cm^2$.
$D_b$ and $D_n$ are the corresponding diffusion coefficients in units of $cm^2/sec$. 
$E_b$ is the bacterial reproduction rate in units of $sec^{-1}$ per nutrition concentration.
$E_n$ is the nutrition consumption rate in units of $sec^{-1}$ per bacteria concentration. 

We change the variables to be dimensionless.
\begin{equation}
\begin{array}  {lr}
\label{dimension}
t\rightarrow tT &
x\rightarrow xX \\
b\rightarrow bB &
n\rightarrow nN 
\end{array}
\end{equation}
the new variables are dimensionless and the capitals are their corresponding units.
We let the temporal and spatial units be the natural scales.
The (microscopic) time scale of the model is the bacterial reproduction time $\tau_R$,
so $T=\tau_{R}$.
The length scale is the diffusion length of the nutrition during reproduction time.
The nutrition available for bacteria during reproduction time 
is proportional to the square of the diffusion length.
The length unit is:
\begin{equation}
\label{length_scale}
X=\sqrt{D_n\tau_R}
\end{equation}
After inserting the dimensionless variables into Eq. (\ref{start}) we obtain:
\begin{eqnarray} 
\partderiv{b}{t}&=&D_b/D_n\nabla^2b+(TN)E_bbn \\
\partderiv{n}{t}&=&\nabla^2n-(TB)E_nbn \nonumber
\end{eqnarray}
We define the relative diffusion coefficient $D\equiv D_b/D_n$ and impose
 the following relations:
\begin{eqnarray}
\label{relation1}
TBE_n&=&1  \\
\label{relation2}
TNE_b&=&1
\end{eqnarray}
We obtain the following dimensionless equations:
\begin{eqnarray} 
\label{eqless}
\partderiv{b}{t}&=&D\nabla^2b+bn \\
\partderiv{n}{t}&=&\nabla^2n-bn \nonumber
\end{eqnarray}

\subsection{Evaluation of the parameters}
\label{evaluation}
We will estimate the values of the parameters $E_b, E_n \mbox{ and } D_n$, 
and derive from them the dimensional units. 
A review of some of the following biological arguments appears in section (\ref{sec:eating}).
\begin{itemize}
\item The bacterial reproduction time, when bacteria grow under optimal conditions,
  is about $\tau_R=25min$.
  Colonies which exhibit branching patterns grow under limited nutrition supply.
  Therefore the reproduction time will be longer, but in the same order of magnitude.
  We set the time unit to be:
  \begin{equation}
    T=\tau_R=25min
  \end{equation}
\item A typical value for the diffusion coefficient of chemicals in agar
  is $10^{-7} cm^2/sec$. So we assume that $D_n$, the diffusion coefficient
  of the nutrition in the agar, is similar. We can find the length unit
  using (\ref{length_scale}):
  \begin{equation}
    X=0.01cm=100\mu m
  \end{equation}
\item The nutrition concentration in the experiments conducted by 
  Ben-Jacob \etal \cite{BenJacob97}, Rafols \cite{Rafols98} and others
  was $0.1-5mg/cm^2$. We set $N$ to have a similar value:
  \begin{equation}
    \label{Nvalue}
    N=1mg/cm^2=10 \times 10^{-12}g/\mu^2=10^{-7}g/X^2
  \end{equation}
\item  The reproduction rate per bacterium (Eq. \ref{start}) is $E_bNn$, where $n$ is the
  dimensionless concentration.
  The rate is the inverse of the reproduction time, which depends on the nutrition concentration.
  We assume that $N$ is the concentration for which the reproduction time is $\tau_R$.
  Therefore:
  \begin{equation}
    \label{growthrate}
    E_bN=1/\tau_R
  \end{equation}
  and Eq. \ref{relation2} is satisfied.

\item Similarly, $E_nN$ is the nutrition consumption rate per bacterium. 
  We suggest that during reproduction time,
  a single bacterium consumes an amount of nutrition three times its mass, 
  which is about $3\times 10^{-12} g$. 
  Therefore, the rate of nutrition consumption is:
  \begin{equation}
    \label{En}
    E_nN\sim\frac{3\times 10^{-12}g\mbox{ nutrition}}{bacteria}\frac{1}{25min}
  \end{equation}
  From (\ref{relation1},\ref{Nvalue}) we obtain the bacterial concentration:
  \begin{equation}
    \label{Bvalue}
    B=N/TE_nN=3bacteria/\mu^2=3\times 10^4 bacteria/X^2
  \end{equation}
\end{itemize}
The discrete time step of the numerical integration is measured in units of $T$.
Sometimes numeric stability demands that the time step will be less than one. 
Then, as an example,
a time step of $0.001$ will correspond to $0.001T\sim 1sec$ 

\subsection{Equations with cutoff in the reaction term}

The scaling performed in the previous section is also 
suitable for the cutoff equations (section \ref{sec:cutoff}). 
We are interested in the meaning of the 
cutoff $\beta$. Since $\beta$ is measured in units of B, we use (\ref{Bvalue}) 
to translate numerical values to bacteria concentrations.
For example, $\beta=3\times 10^{-5}$ corresponds to $1bacteria/X^2$,
while $\beta=1$ corresponds approximately to bacteria covering 
the agar surface in a continuous layer.

\subsection{Mimura \etal model}

The model's equations in dimensional form (section \ref{metastable}) are:
\begin{eqnarray} 
\partderiv{b}{t}&=&D_b\nabla^2b+E_bbn-\frac{a_0b}{(a_n+n)(a_b+b)} \\
\partderiv{n}{t}&=&D_n\nabla^2n-E_nbn \nonumber\\
\partderiv{s}{t}&=&\frac{a_0b}{(a_n+n)(a_b+b)} \nonumber
\end{eqnarray}
where $a_n$ and $a_b$ are constants which have the same units as $n$ and $b$ respectively.
We introduce dimensionless variables as previously (Eq. \ref{dimension}), and 
impose relation (\ref{relation1}). We obtain:
\begin{eqnarray} 
\partderiv{b}{t}&=&D_b/D_n\nabla^2b
+\frac{E_bN}{E_nB}bn-\frac{a_0T}{BN}\frac{b}{(a_n/N+n)(a_b/B+b)} \\
\partderiv{n}{t}&=&\nabla^2n-bn \nonumber\\
\partderiv{s}{t}&=& \frac{a_0T}{BN}\frac{b}{(a_n/N+n)(a_b/B+b)} \nonumber
\end{eqnarray}
We define new parameters:
\begin{eqnarray}
D&\equiv&D_b/D_n \\
\label{eps}
\epsilon&\equiv&\frac{E_bN}{E_nB} \\
\mu&\equiv&\frac{a_0T}{BN} 
\end{eqnarray}
We set $a_n/N=1$ and $a_b/B=1$ in the last term of the bacterial equation.
(The assignment is acceptable since Mimura \etal does not justify the exact form
of that term, rather they state that it is only its general properties that matters.)
Changing to the new parameters gives the dimensionless equations.

Scaling this model requires adjusting the values of the units. Relation (\ref{relation2})
is replaced by relation (\ref{eps}), which is equivalent to:
\begin{equation}
\label{Endelta}
TNE_b=\epsilon
\end{equation}
Since (\ref{growthrate}) is still valid, we leave $N$ unchanged. The other units re-scale 
according to:
\begin{eqnarray} 
T&\rightarrow&\epsilon T \\
X&\rightarrow&\sqrt{\epsilon} X \nonumber\\
B&\rightarrow&B/\epsilon \nonumber
\end{eqnarray}
compared to the values of the units evaluated in section (\ref{evaluation})

%


\begin{thebibliography}{10}

\bibitem{SDA57}
M.~Doudoroff R.~Y.~Stainer and E.~A. Adelberg.
\newblock {\em The Microbial World}.
\newblock Prentice-Hall and Inc., N. J., 1957.

\bibitem{Shap88}
J.~A. Shapiro.
\newblock Bacteria as multicellular organisms.
\newblock {\em Sci. Am.}, 258(6):62--69, 1988.

\bibitem{BenJacob97}
E.~{Ben-Jacob}.
\newblock From snowflake formation to the growth of bacterial colonies. part
  {II}: Cooperative formation of complex colonial patterns.
\newblock {\em Contemp. Phys.}, 38:205--241, 1997.

\bibitem{BCL98}
E.~{Ben-Jacob}, I.~Cohen, and H.~Levine.
\newblock Cooperative self-organization of microorganisms.
\newblock {\em Adv. Phys.}, 1998.
\newblock (in press).

\bibitem{LB98}
H.~Levine and E.~{Ben-Jacob}.
\newblock The art and science of microorganisms.
\newblock {\em Sci. Am.}, 1998.
\newblock (in press).

\bibitem{Mend78}
N.~H. Mendelson.
\newblock Helical {{\bsub*}} macrofibers: morphogenesis of a bacterial
  multicellular macroorganism.
\newblock {\em Proc. Natl. Acad. Sci. USA}, 75(5):2478--2482, 1978.

\bibitem{Devreotes89}
P.~Devreotes.
\newblock {\it Dictyostelium discoideum}: a model system for cell-cell
  interactions in development.
\newblock {\em Science}, 245:1054--1058, 1989.

\bibitem{MKNIHY92}
Matsuyama T; Kaneda K; Nakagawa Y; Isa K; H. Hara-Hotta;~Yano I.
\newblock A novel extracellular cyclic lipopeptide which promotes
  flagellum-dependent and -independent spreading growth of {\it serratia
  marcescens}.
\newblock {\em J. Bacteriol.}, 174:1769--1776, 1992.

\bibitem{Harshey94}
R.~M. Harshey.
\newblock Bees aren't the only ones: swarming in gram-negative bacteria.
\newblock {\em Molecular Microbiology}, 13:389--394, 1994.

\bibitem{FWG94}
Fuqua WC; Winans SC;~Greenberg EP.
\newblock Quorum sensing in bacteria: the {LuxR-LuxI} family of cell
  density-responsive transcriptional regulators.
\newblock {\em J. Bacteriol.}, 176:269--275, 1994.

\bibitem{LWFBSLW95}
A.~Latifi, M.~K. Winson, M.~Foglino, B.~W. Bycroft, G.~S. Stewart,
  A.~Lazdunski, and P.~Williams.
\newblock Multiple homologues of luxr and luxi control expression of virulence
  determinants and secondary metabolites through quorum sensing in {\it
  pseudomonas aeruginosa} pao1.
\newblock {\em Molecular Microbiology}, 17:333--343, 1995.

\bibitem{FWG96}
C.~Fuqua, S.~C. Winans, and E.~P. Greenberg.
\newblock Census and consensus in bacterial ecosystems: the {LuxR-LuxI} family
  of quorum-sensing transcriptional regulators.
\newblock {\em Annu. Rev. Microbiol.}, 50:727--751, 1996.

\bibitem{BB91}
E.~O. Budrene and H.~C. Berg.
\newblock Complex patterns formed by motile cells of {\em esherichia coli}.
\newblock {\em Nature}, 349:630--633, 1991.

\bibitem{BE95}
Y.~Blat and M.~Eisenbach.
\newblock Tar-dependent and -independent pattern formation by {\salmon*}.
\newblock {\em J. Bac.}, 177(7):1683--1691, 1995.

\bibitem{BB95}
E.~O. Budrene and H.~C. Berg.
\newblock Dynamics of formation of symmetrical patterns by chemotactic
  bacteria.
\newblock {\em Nature}, 376:49--53, 1995.

\bibitem{ST91}
J.~A. Shapiro and D.~Trubatch.
\newblock Sequential events in bacterial colony morphogenesis.
\newblock {\em Physica D}, 49:214--223, 1991.

\bibitem{SM93}
B.~Salhi and N.~H. Mendelson.
\newblock Patterns of gene expression in {\bsub*} colonies.
\newblock {\em J. Bacteriol.}, 175:5000--5008, 1993.

\bibitem{MS96}
N.~H. Mendelson and B.~Salhi.
\newblock Patterns of reporter gene expression in the phase diagram of {\bsub*}
  colony forms.
\newblock {\em J. Bacteriol.}, 178:1980--1989, 1996.

\bibitem{GR95}
T.~Galitski and J.~R. Roth.
\newblock Evidence that {F} plasmid transfer replication underlies apparent
  adaptive mutation.
\newblock {\em Science}, 268:421--423, 1995.

\bibitem{RPF95}
J.~P. Rasicella, P.~U. Park, and M.~S. Fox.
\newblock Adaptive mutation in {\ecoli*} : a role for conjugation.
\newblock {\em Science}, 268:418--420, 1995.

\bibitem{Miller98}
R.~V. Miller.
\newblock Bacterial gene swapping in nature.
\newblock {\em Sci. Am.}, January 1998, 1998.

\bibitem{Kessler85}
J.~O. Kessler.
\newblock Co-operative and concentrative phenomena of swimming micro-organisms.
\newblock {\em Cont. Phys.}, 26:147--166, 1985.

\bibitem{FM89}
H.~Fujikawa and M.~Matsushita.
\newblock Fractal growth of {\bsub*} on agar plates.
\newblock {\em J. Phys. Soc. Jap.}, 58:3875--3878, 1989.

\bibitem{PK92a}
T.~J. Pedley and J.~O. Kessler.
\newblock Bioconvection.
\newblock {\em Sci. Prog.}, 76:105--123, 1989.

\bibitem{BSST92}
E.~{Ben-Jacob}, H.~Shmueli, O.~Shochet, and A.~Tenenbaum.
\newblock Adaptive self-organization during growth of bacterial colonies.
\newblock {\em Physica A}, 187:378--424, 1992.

\bibitem{MHM93}
T.~Matsuyama, R.~M. Harshey, and M.~Matsushita.
\newblock Self-similar colony morphogenesis by bacteria as the experimental
  model of fractal growth by a cell population.
\newblock {\em Fractals}, 1(3):302--311, 1993.

\bibitem{BSTCCV94a}
E.~{Ben-Jacob}, O.~Shochet, A.~Tenenbaum, I.~Cohen, A.~Czir\'ok, and T.~Vicsek.
\newblock Generic modeling of cooperative growth patterns in bacterial
  colonies.
\newblock {\em Nature}, 368:46--49, 1994.

\bibitem{BCSALT95}
E.~{Ben-Jacob}, I.~Cohen, O.~Shochet, I.~Aronson, H.~Levine, and L.~Tsimering.
\newblock Complex bacterial patterns.
\newblock {\em Nature}, 373:566--567, 1995.

\bibitem{WTMMBB95}
D.~E. Woodward, R.~Tyson, M.~R. Myerscough, J.~D. Murray, E.~O. Budrene, and
  H.~C. Berg.
\newblock Spatio-temporal patterns generated by salmonella typhimurium.
\newblock {\em Biophys. J.}, 68:2181--2189, 1995.

\bibitem{BCCVG97}
E.~{Ben-Jacob}, I.~Cohen, A.~Czir\'ok, T.~Vicsek, and D.~L. Gutnick.
\newblock Chemomodulation of cellular movement and collective formation of
  vortices by swarming bacteria and colonial development.
\newblock {\em Physica A}, 238:181--197, 1997.

\bibitem{KW97}
J.~O. Kessler and M.~F. Wojciechowski.
\newblock Collective behavior and dynamics of swimming bacteria.
\newblock In J.~A. Shapiro and M.~Dworkin, editors, {\em Bacteria as
  Mullticellular Organisms}, pages 417--450. Oxford University Press Inc., New
  York, 1997.

\bibitem{ES98}
S.~E. Esipov and J.~A. shapiro.
\newblock Kinetic model of {\em proteus mirabilis} swarm colony development.
\newblock {\em J. Math. Biol.}, 36:249--268, 1998.

\bibitem{MF90}
M.~Matsushita and H.~Fujikawa.
\newblock Diffusion-limited growth in bacterial colony formation.
\newblock {\em Physica A}, 168:498--506, 1990.

\bibitem{FM91}
H.~Fujikawa and M.~Matsushita.
\newblock Bacterial fractal growth in the concentration field of nutrient.
\newblock {\em J. Phys. Soc. Jap.}, 60:88--94, 1991.

\bibitem{SC38}
R.~N. Smith and F.~E. Clacrk.
\newblock Motile colonies of bacillus alvei and other bacteria.
\newblock {\em J. Bacteriol.}, 35:59--60, 1938.

\bibitem{Henrici48}
T.~H. Henrici.
\newblock {\em The Biology of Bacteria: The Bacillaceae}.
\newblock D. C. Heath \& company, 3rd edition, 1948.

\bibitem{WitSan81}
T.~A. Witten and L.~M. Sander.
\newblock Diffusion--limited aggregation, a kinetic critical phenomenon.
\newblock {\em Phys. Rev. Lett.}, 47:1400, 1981.

\bibitem{Sander86}
L.M. Sander.
\newblock Fractal growth processes.
\newblock {\em Nature}, 322:789--793, 1986.

\bibitem{Vicsek89}
T.~Vicsek.
\newblock {\em Fractal Growth Phenomena}.
\newblock World Scientific, New York, 1989.

\bibitem{MM95}
T.~Matsuyama and M.~Matsushita.
\newblock Morphogenesis by bacterial cells.
\newblock In P.~M. Iannaccone and M.~K. Khokha, editors, {\em Farctal Geometry
  in Biological Systems, an Analytical Approach}, pages 127--171. CRC Press,
  New-York, 1995.

\bibitem{BTSA94}
E.~{Ben-Jacob}, A.~Tenenbaum, O.~Shochet, and O.~Avidan.
\newblock Holotransformations of bacterial colonies and genome cybernetics.
\newblock {\em Physica A}, 202:1--47, 1994.

\bibitem{TBG98}
M.~Tcherpikov, E.~{Ben-Jacob}, and D.~Gutnick.
\newblock Identification of two pattern-forming strains and their localization
  in a phylogenetic cluster.
\newblock {\em Int. J. Syst. Bacteriol.}, 1998.
\newblock (in press).

\bibitem{BCG98}
E.~{Ben-Jacob}, I.~Cohen, and D.~Gutnick.
\newblock Cooperative organization of bacterial colonies: From genotype to
  morphotype.
\newblock {\em Annu. Rev. Microbiol.}, 1998.
\newblock (in press).

\bibitem{MM93}
T.~Matsuyama and M.~Matsushita.
\newblock Fractal morphogenesis by a bacterial cell population.
\newblock {\em Crit. Rev. Microbiol.}, 19:117--135, 1993.

\bibitem{KKL88}
D.~A. Kessler, J.~Koplik, and H.~Levine.
\newblock Pattern selection in fingered growth phenomena.
\newblock {\em Adv. Phys.}, 37:255, 1988.

\bibitem{Langer89}
J.S. Langer.
\newblock Dendrites, viscous fingering, and the theory of pattern formation.
\newblock {\em Science}, 243:1150--1154, 1989.

\bibitem{BG90}
E.~{Ben-Jacob} and P.~Garik.
\newblock The formation of patterns in non-equilibrium growth.
\newblock {\em Nature}, 343:523--530, 1990.

\bibitem{BenJacob93}
E.~{Ben-Jacob}.
\newblock From snowflake formation to the growth of bacterial colonies. part
  {I}: Diffusive patterning in non-living systems.
\newblock {\em Contemp. Phys.}, 34:247--273, 1993.

\bibitem{Horgan95}
J.~Horgan.
\newblock From complexity to perplexity.
\newblock {\em Sci. Am.}, June 95:74--79, 1995.

\bibitem{KL93}
D.~A. Kessler and H.~Levine.
\newblock Pattern formation in {\em dictyostelium} via the dynamics of
  cooperative biological entities.
\newblock {\em Phys. Rev. E}, 48:4801--4804, 1993.

\bibitem{Azbel93}
M.~Y. Azbel.
\newblock Survival-extinction transition in bacteria growth.
\newblock {\em Europhys. Lett.}, 22(4):311--316, 1993.

\bibitem{BCSCV95}
E.~{Ben-Jacob}, I.~Cohen, O.~Shochet, A.~Czir\'ok, and T.~Vicsek.
\newblock Cooperative formation of chiral patterns during growth of bacterial
  colonies.
\newblock {\em Phys. Rev. Lett.}, 75(15):2899--2902, 1995.

\bibitem{KLT97}
D.~A. Kessler, H.~Levine, and L.~Tsimring.
\newblock Computational modeling of mound development in {\em dictyostelium}.
\newblock {\em Physica D}, 106(3-4):375--388, 1997.

\bibitem{PS78}
H.~Parnas and L.~Segel.
\newblock A computer simulation of pulsatile aggregation in {{\it Dictyostelium
  discoideum}}.
\newblock {\em J. Theor. Biol.}, 71:185--207, 1978.

\bibitem{Mackay78}
S.~A. Mackay.
\newblock Computer simulation of aggregation in dictyostelium discoideum.
\newblock {\em J. Cell. Sci.}, 33:1--16, 1978.

\bibitem{Mimura97}
M.~Mimura, H.~Sakaguchi, and M.~Matsushita.
\newblock A reaction-diffusion approach to bacterial colony formation.
\newblock {\em preprint}, 1997.

\bibitem{MWIRMSM98}
M.~Matsushita, J.~Wakita, H.~Itoh, I.~Rafols, T.~Matsuyama, H.~Sakaguchi, and
  M.~Mimura.
\newblock Interface growth and pattern formation in bacterial colonies.
\newblock {\em Physica A}, 249:517--524, 1998.

\bibitem{KMMUS97}
K.~Kawasaki, A.~Mochizuki, M.~Matsushita, T.~Umeda, and N.~Shigesada.
\newblock Modeling spatio-temporal patterns created by bacillus-subtilis.
\newblock {\em J. Theor. Biol.}, 188:177--185, 1997.

\bibitem{Kitsunezaki97}
S.~Kitsunezaki.
\newblock Interface dynamics for bacterial colony formation.
\newblock {\em J. Phys. Soc. Jpn}, 66(5):1544--1550, 1997.

\bibitem{Rafols98}
I. Rafols, {\em Formation of concentric rings in bacterial colonies}, MSc
  thesis, Chuo University, Japan, 1998.

\bibitem{KL98}
D. A. Kessler and H. Levine,''Fluctuation-induced diffusive instabilities'',
  Nature, in press.

\bibitem{BSTCCV94b}
E.~{Ben-Jacob}, O.~Shochet, A.~Tenenbaum, I.~Cohen, A.~Czir\'ok, and T.~Vicsek.
\newblock Communication, regulation and control during complex patterning of
  bacterial colonies.
\newblock {\em Fractals}, 2(1):15--44, 1994.

\bibitem{BJGMG88}
E.~{Ben-Jacob}, P.~Garik, T.~Muller, and D.~Grier.
\newblock Characterization of morphology transitions in diffusion-controlled
  systems.
\newblock {\em Phys. Rev. A}, 38:1370, 1988.

\bibitem{Eis90}
M.~Eisenbach.
\newblock Functions of the flagellar modes of rotation in bacterial motility
  and chemotaxis.
\newblock {\em Molec. Microbiol.}, 4(2):161--167, 1990.

\bibitem{Henri72}
J.~Henrichsen.
\newblock Bacterial surface translocation: a survey and a classification.
\newblock {\em Bac. Rev.}, 36:478--503, 1972.

\bibitem{Berg93}
H.~C. Berg.
\newblock {\em Random Walks in Biology}.
\newblock Princeton University Press, Princeton, N.J., 1993.

\bibitem{Adler69}
J.~Adler.
\newblock Chemoreceptors in bacteria.
\newblock {\em Science}, 166:1588--1597, 1969.

\bibitem{BP77}
H.~C. Berg and E.~M. Purcell.
\newblock Physics of chemoreception.
\newblock {\em Biophysical Journal}, 20:193--219, 1977.

\bibitem{Lacki81}
J.~M. Lackiie, editor.
\newblock {\em Biology of the chemotactic response}.
\newblock Cambridge Univ. Press, 1986.

\bibitem{Dusenbery98}
D.~B. Dusenbery.
\newblock Spatial sensing of stimulus gradients can be superior to temporal
  sensing for free-swimming bacteria.
\newblock {\em Biophys. J.}, 74:2272--2277, 1998.

\bibitem{Adler66}
J.~Adler.
\newblock Chemotaxis in bacteria.
\newblock {\em Science}, 153:708--716, 1966.

\bibitem{BB72}
H.~C. Berg and D.~A. Brown.
\newblock Chemotaxis in {\ecoli*} analyzed by three-dimensional tracking.
\newblock {\em Nature}, 239:500--504, 1972.

\bibitem{Murray89}
J.~D. Murray.
\newblock {\em Mathematical Biology}.
\newblock Springer-Verlag, Berlin, 1989.

\bibitem{WB89}
A.~J. Wolfe and H.~C. Berg.
\newblock Migration of bacteria in semisolid agar.
\newblock {\em Proc. Natl. Acad. Sci. USA}, 86:6973--6977, 1989.

\bibitem{CCB96}
I.~Cohen, A.~Czir\'ok, and E.~{Ben-Jacob}.
\newblock Chemotactic-based adaptive self organization during colonial
  development.
\newblock {\em Physica A}, 233:678--698, 1996.

\bibitem{Fisher37}
R.~A. Fisher.
\newblock The wave of advance of advantageous genes.
\newblock {\em Annual Eugenics}, 7:255--369, 1937.

\bibitem{KPP37}
A.~I. Kolmogorov, I.~Petrovsky, and N.~Piscounov.
\newblock \'etude de l'\'equation de la diffusion avec croissance de la
  quantit\'e de mati\`ere et son application \`a un probl\`eme biologique.
\newblock {\em Moscow Univ. Bull. Math.}, 1:1--25, 1937.

\bibitem{AW75}
D.~G. Aronson and H.~F. Weinberger.
\newblock Nonlinear diffusion in population genetics, combustion, and nerve
  pulse propagation.
\newblock In {\em Partial Differential Equations and Related Topics}, Lecture
  Notes in Mathematics, pages 5--49. Springer-Verlag, Berlin, 1975.

\bibitem{BBDKL85}
E.~Ben-Jacob, H.~Brand, G.~Dee, L.~Kramer, and J.S. Langer.
\newblock Pattern propagation in nonlinear dissipative systems.
\newblock {\em Physica D}, 14:348--364, 1985.

\bibitem{vanSaa88}
W.~van Saarloos.
\newblock Front propagation into unstable states: marginal stability as a
  dynamical mechanism for velocity selection.
\newblock {\em Phys. Rev. A}, 37:211--229, 1988.

\bibitem{vanSaa89}
W.~van Saarloos.
\newblock Front propagation into unstable states: {II}. linear versus
  non-linear marginal stability and rate of convergence.
\newblock {\em Phys. Rev. A}, 39:6367--6389, 1989.

\bibitem{cross}
M.Cross, P.Hohenberg, Rev. Mod. Phys. {\bf 65}, 851(1993).

\bibitem{KNS98}
D. A. Kessler, Z. Ner and L. M. Sander, ``Front propagation: Precursors,
  cutoffs and structural stability'', preprint 1998.

\bibitem{MS64}
W.~W. Mullins and R.~F. Sekerka.
\newblock Stability of a planar interface during solidification of a dilute
  binary alloy.
\newblock {\em J. Appl. Phys.}, 35:444, 1964.

\bibitem{MWM95}
M.~Matsushita, J.-I. Wakita, and T.~Matsuyama.
\newblock Growth and morphological changes of bacteria colonies.
\newblock In P.~E. Cladis and P.~{Palffy-Muhoray}, editors, {\em
  Spatio-Temporal Patterns in Nonequilibrium Complex Systems}, {Santa-Fe}
  Institute studies in the sciences of complexity, pages 609--618.
  Addison-Weseley Publishing Company, 1995.

\bibitem{BSTA95}
E.~{Ben-Jacob}, O.~Shochet, A.~Tenenbaum, and O.~Avidan.
\newblock Evolution of complexity during growth of bacterial colonies.
\newblock In P.~E. Cladis and P.~{Palffy-Muhoray}, editors, {\em
  Spatio-Temporal Patterns in Nonequilibrium Complex Systems}, {Santa-Fe}
  Institute studies in the sciences of complexity, pages 619--634.
  Addison-Weseley Publishing Company, 1995.

\bibitem{WKNM94}
J.~Wakita, K.~Komatsu, A.~Nakahara, T.~Matsuyama, and {\it et al.}
\newblock Experimental investigation on the validity of population dynamics
  approach to bacterial colony formation.
\newblock {\em J Physical Soc. Japan}, 63:1205--1211, 1994.

\bibitem{KSB94}
R.~Kupferman, O.~Shochet, and E.~{Ben-Jacob}.
\newblock Numerical study of morphology diagram in the large undercooling limit
  using a phase-field model.
\newblock {\em Phys. Rev. E}, 50:1005, 1994.

\bibitem{Kupferman95}
R.~Kupferman.
\newblock {\em Morphology, coexistence and selection in interfacial pattern
  formation}.
\newblock PhD thesis, Tel-Aviv University, 1995.

\bibitem{Cohen97}
I. Cohen, {\em Mathematical Modeling and Analysis of Pattern Formation and
  Colonial Organization in Bacterial Colonies}, MSc thesis, Tel-Aviv
  University, ISRAEL, 1997.

\bibitem{BLC98}
E. Ben-Jacob, H. Levine and I. Cohen,,''Cooperative self-organization of
  microorganisms'', to appear in Adv. Phys.

\bibitem{Newman83}
W.~I. Newman.
\newblock The long-time behavior of the solution to a non-linear diffusion
  problem in population genetics and combustion.
\newblock {\em J. Theor. Biol.}, 104:473--484, 1983.

\bibitem{TLR93}
Y.~Tu, H.~Levine, and D.~Ridgway.
\newblock Morphology transition in a mean-field model of diffusion-limited
  growth.
\newblock {\em Phys. Rev. Lett.}, 71(23):3838--3841, 1993.

\bibitem{BSCTCV95}
E.~Ben-Jacob, O.~Shochet, I.~Cohen, A.~Tenenbaum, A.~Czir\'ok, and T.~Vicsek.
\newblock Cooperative strategies in formation of complex bacterial patterns.
\newblock {\em Fractals}, 3:849--868, 1995.

\end{thebibliography}


\newpage

\begin{figure}
\caption {
\label{fig:T:example}
Typical example of branching growth of the \T morphotype for $1g/l$
peptone level and 1.5\% agar concentration.
}
\end{figure}

\begin{figure}
\caption {
\label{fig:T:morphology}
Examples of typical  patterns of \Tm for intermediate agar
concentration (clockwise from top left).
(a) At very high peptone level (peptone 12$g/l$, agar concentration 1.75\%)
the pattern is compact.
(b) At high peptone level (3$g/l$, agar 2\%) the pattern is of dense
fingers with pronounced radial symmetry -- similar to patterns
observed in Hele-Show cell.
(c) At intermediate peptone level (1$g/l$, agar 1.75\%) the pattern is
"bushy" fractal-like pattern, with branch width smaller than the
distance between branches.
(d) At low peptone level (0.1$g/l$, agar 1.75\%) there are fine radial
branches with apparent circular envelope.
}
\end{figure}

\begin{figure}
\caption {
\label{fig:T:extremes}
(left) Fractal pattern for $0.01g/l$ peptone level and 1.75\%
agar concentration.
(right) Dense branching pattern for $4g/l$ peptone and 2.5\% agar.
Note that the branches are much thinner than those in Fig. 2b, i.e.
the branches are thinner for higher agar concentrations.
}
\end{figure}

\begin{figure}
\caption {
\label{fig:T:fricks}
Patterns of colonies of \T morphotype (clockwise from top left). 
(a) Pattern of concentric rings superimposed on a branched colony for
2.5$g/l$ peptone level and 2.5\% agar concentration.
(b) Concentric rings in a compact growth for $15g/l$ peptone level and
2.25\% agar concentration.
(c) Weak chirality (global twist of the branches) for $4g/l$ peptone
and 2.5\% agar concentration.
(d) Closer look at the branches ($\times50$ magnification) shows
density variations within each branch.
Darker colors represent thicker layer of bacteria.
}
\end{figure}

\begin{figure}
\caption {
\label{fig:T:transition}
Coexistence of two morphologies near the critical point. The colony
shows combination of the radial symmetry morphology and the
fractal-like morphology. The colony is grown on agar concentration of
1.75\% and 1$g/l$ peptone level.
}
\end{figure}

\begin{figure}
\caption {
\label{fig:bsub:morphology}
Morphology diagram of \bsub colonies grown by Matsushita \etal
\protect\cite{FM89,MF90,MHM93}. Taken with permission from
\protect\cite{Rafols98}.
}
\end{figure}

\begin{figure}
\caption {
\label{fig:bsub:rings}
Colony of \bsub-. Pattern of concentric rings superimposed on a
branched colony. Taken with permission from
\protect\cite{Rafols98}.
}
\end{figure}

\begin{figure}
\caption {
\label{fig:spores}
Electron microscope observation of \T bacteria. Round or oval shapes
with bright center are spores. Elongated shapes are living cells.
The cells engolfing oval shapes are pre-spores.
}
\end{figure}

\begin{figure}[]
\caption [A Picture]
        {\protect\footnotesize Reaction term $f(b)$ (above) and
          Landau-Ginzburg free energy $\Phi(b)$ (below) for the
          Fisher-Kolmogorove equation.} 
\label{fisher-fig}
\end{figure}


\begin{figure}[]
\caption[]
{\protect\footnotesize Typical front obtained for the 1D diffusive
  Fisher-Kolmogorove model. Parameters: $D_b=0.1, D_n=1$, Initial food
  concentration $n_0=1$.} 
\label{fisher-front-fig}
\end{figure}


\begin{figure}[]
\caption[]
{\protect\footnotesize  Landau-Ginzburg free
          energy $\Phi(b)$ for the solid-liquid phase-field
          model. Note the meta-stable point at $b=0$ and the stable
          one at $b=1$.} 
\label{meta-fig}
\end{figure}


\begin{figure}[]
\caption[]
{\protect\footnotesize Growth term $f(b)$ (above) and Landau-Ginzburg free
  energy $\Phi(b)$ (below) for
  the Kessler-Levine correction. Cutoff value $\beta=1$, nutrient level $n=1$.} 
\label{kl-fig}
\end{figure}


\begin{figure}[]
\caption[]
{\protect\footnotesize 2D growth pattern of the Kessler-Levine correction
  with no death term. Parameters are: $Db=0.01, \beta=0.25, n_0=1$}
\label{kl-pattern-fig}
\end{figure}


\begin{figure}[]
\caption[]
{\protect\footnotesize 2D growth pattern $(b+s)$ of the Kessler-Levine
  correction with a death term. Parameters are:  $Db=0.01, \beta=0.25,
  n_0=1, \mu=0.01$.}
\label{kl-death-fig}
\end{figure}



\begin{figure}[]
\caption[]
{\protect\footnotesize 
  2D growth patterns $(b+s)$ of the
  model with lubrication.
  a. Parameters are: $\gamma=-1/2, \lambda=0.01, \eta=2, D_b=0.5,
       D_l=0.5, \mu=1, n_0=1.5, j=0.01,
       l_{max}=2, \Gamma=1$ 
  b. Parameters are as in Fig. a, except a higher
     absorption rate $\lambda=0.1$.
  c. Parameters are as in Fig. a, except a different exponent for 
     the bacterial diffusion $\gamma=-1$.
  d. Parameters are as in Fig. c, except a higher
     absorption rate $\lambda=0.03$ 
  }
\label{yony-fig}
\end{figure}

\begin{figure}[]
\caption[]
{\protect\footnotesize
  Profile of the fronts of the bacterial field (solid line)
  and of the lubricating field (dashed line) from the 2D
  model with lubrication.
  The bacterial field was scaled by a factor of 40.
  left: Parameters are as in Fig.\ \ref{yony-fig}a.
  right: Parameters are as in Fig.\ \ref{yony-fig}c.
}
\label{yony-front}
\end{figure}

\begin{figure}[]
\caption[]
{\protect\footnotesize 1D front obtained for the Kitsunezaki model. 
Parameters are: $D_0=0.1, k=1, \mu=0.15, n_0=1$}
\label{kits-k1-fig}
\end{figure}


\begin{figure}[]
\caption[]
{\protect\footnotesize 1D front obtained for the Kitsunezaki model
  with k=2. All other parameters as in Fig.\ \ref{kits-k1-fig}.}
\label{kitz-k2-fig}
\end{figure}


\begin{figure}[]
\caption[]
{\protect\footnotesize 2D growth pattern $(b+s)$ of the Kitsunezaki
  model. Parameters are:  $D_0=0.1, k=1, \mu=0.15, n_0=1$.}
\label{kitz-pattern-fig}
\end{figure}


\begin{figure}[]
\caption[]
{\protect\footnotesize 2D growth pattern $(b)$ of the Kawasaki \etal
  model.
  Parameters are: $D_b = n*b, n_0=0.71$.} 
\label{kawa-pattern-fig}
\end{figure}


\begin{figure}[]
\caption[]
{\protect\footnotesize 
  Growth term $f(b)$ (above) and Landau-Ginzburg free
  energy $\Phi(b)$ (below) for the Mimura \etal model.
  Parameters are: $\epsilon=20,
  \mu=2400, n=5$. Note the stable point at $b=0$.}
\label{mimura-fig}
\end{figure}


\begin{figure}[]
\caption[]
{\protect\footnotesize  
  1D front obtained for the Mimura \etal  model.
  Parameters are: $D_b=0.1, n_0=10, \epsilon=20, \mu=2400$.} 
\label{mimura-front-fig}
\end{figure}


\begin{figure}[]
\caption[]
{\protect\footnotesize Various 2D patterns $(b+s)$ obtained for the
  Mimura \etal model:
  a. DLA-like $(D_b=0.05, n_0=10)$.
  b. dense branches $(D_b=0.09, n_0=10)$.
  b. Concentric rings $(D_b=0.05, n_0=12)$.
  c. Compact growth $(D_b=0.1, n_0=14)$.
  In all cases, $\mu=2400, \epsilon=20$.}
\label{mimura-patterns-fig1}
\end{figure}


\begin{figure}[]
\caption[]
{\protect\footnotesize Mimura \etal model: The effect of changing the
  initial nutrient level $n_0$. For all pictures $D_b=0.05, \mu=2400, \epsilon=20$.}
\label{mimura-patterns-fig2}
\end{figure}


\begin{figure}[]
\caption[]
{\protect\footnotesize Mimura \etal model: The effect of changing
  $D_b$. 
  For all pictures $n_0=10, \mu=2400, \epsilon=20$.}
\label{mimura-patterns-fig3}
\end{figure}


\begin{figure}[]
\caption[]
{\protect\footnotesize Colonial growth velocity vs. $D_b$ for the
  Mimura \etal model. Parameter values as in Fig.\ \ref{mimura-patterns-fig3}.}
\label{mimura-vel-fig}
\end{figure}



\begin{figure}[]
\caption[]
{\protect\footnotesize  
  2D growth pattern $(b+s)$ of the Kitsunezaki
  model with food chemotaxis included.  $\chi_{0f}=3$.
  Other parameters as in Fig.\ \ref{kitz-pattern-fig}.}
\label{kits-food-fig}
\end{figure}

\begin{figure}[]
\caption[]
{\protect\footnotesize 2D growth pattern $(b+s)$ of the Kitsunezaki
  model with repulsive chemotactic signaling included. Parameters
  are: $\chi_{0r}=1, D_r=1, \Gamma_r=0.25, \Omega_r=0, \Lambda_r=0.001$.
  Other parameters as in Fig.\ \ref{kitz-pattern-fig}.}
\label{kits-rep-fig}
\end{figure}


\begin{figure}[]
\caption[]
{\protect\footnotesize 
  2D growth pattern $(b+s)$ of the Mimura \etal
  model with food chemotaxis included.  $\chi_{0f}=0.06$.
  Other parameters as in Fig.\ \ref{mimura-patterns-fig1} a.}
\label{mimura-food-fig}
\end{figure}

\begin{figure}[]
\caption[]
{\protect\footnotesize  
  2D growth pattern $(b+s)$ of the Mimura \etal
  model with repulsive chemotactic signaling included. Parameters
  are: $\chi_{0r}=0.1, D_r=1, \Gamma_r=0.2, \Omega_r=0, \Lambda_r=0.01$.
  Other parameters as in Fig.\ \ref{mimura-patterns-fig1} a.}
\label{mimura-rep-fig}
\end{figure}


\end{document}